\documentclass[12pt]{article}
\usepackage[latin1]{inputenc}

\usepackage{lipsum}
\usepackage{amsmath}
\usepackage{color}
\usepackage{amsfonts}
\usepackage[mathscr]{euscript}
\usepackage{graphicx}
\usepackage{geometry}
\usepackage{amssymb,epsfig}
\usepackage{comment}
\usepackage{tabularx}
\usepackage{bm}
\usepackage[dvipsnames]{xcolor}
\usepackage{exscale}
\usepackage{amsbsy}
\usepackage{textcomp}
\usepackage{hyperref}
\usepackage{slashed}
\usepackage{authblk}
\usepackage{tabularx}
\usepackage{color}
\usepackage{doi}
\usepackage{tensor}
\usepackage{wrapfig}
\usepackage[font=footnotesize,labelfont=bf]{caption}
\usepackage{ulem} 
\usepackage{tikz-feynman}
\usepackage{slashed}
\usepackage{mathrsfs}

\usepackage{makecell}

\def\ba#1\ea{\begin{align}#1\end{align}}
\def\bg#1\eg{\begin{gather}#1\end{gather}}
\def\bm#1\em{\begin{multline}#1\end{multline}}
\def\bmd#1\emd{\begin{multlined}#1\end{multlined}}

\newcommand{\be}{\begin{equation}}
	\newcommand{\ee}{\end{equation}}
\newcommand{\bea}{\begin{eqnarray}}
	\newcommand{\eea}{\end{eqnarray}}

\newcommand{\bs}{\boldsymbol}
\newcommand{\matleft}{\left(\begin{array}}
	\newcommand{\matright}{\end{array}\right)}
\newcommand{\Tr}{\operatorname{Tr}}

\newcommand{\sgn}{\operatorname{sgn}}

\usepackage{subcaption}
\usepackage{bbm}

\usepackage[numbers,sort&compress]{natbib}
\def\simge{
	\mathrel{\rlap{\raise 0.511ex 
			\hbox{$>$}}{\lower 0.511ex \hbox{$\sim$}}}}

\def\simle{
	\mathrel{\rlap{\raise 0.511ex 
			\hbox{$<$}}{\lower 0.511ex \hbox{$\sim$}}}}

\makeatletter
\renewcommand\section{\@startsection {section}{1}{\z@}%
	{-3.5ex \@plus -1ex \@minus -.2ex}
	{2.3ex \@plus.2ex}%
	{\normalfont\large\bfseries}}
\renewcommand\subsection{\@startsection{subsection}{2}{\z@}%
	{-3.25ex\@plus -1ex \@minus -.2ex}%
	{1.5ex \@plus .2ex}%
	{\normalfont\bfseries}}
\renewcommand\subsubsection{\@startsection{subsubsection}{3}{\z@}%
	{-3.25ex\@plus -1ex \@minus -.2ex}%
	{1.5ex \@plus .2ex}%
	{\normalfont\itshape}}
\makeatother

\def\pplogo{\vbox{\kern-\headheight\kern -29pt
		\halign{##&##\hfil\cr&{\ppnumber}\cr\rule{0pt}{2.5ex}&\ppdate\cr}}}
\makeatletter
\def\ps@firstpage{\ps@empty \def\@oddhead{\hss\pplogo}%
	\let\@evenhead\@oddhead 
}
\hypersetup{
	unicode=false,          
	pdftoolbar=true,        
	pdfmenubar=true,        
	pdffitwindow=false,     
	pdfstartview={FitH},    
	pdftitle={anyon thermodynamics},    
	pdfauthor={},     
	pdfsubject={Subject},   
	pdfcreator={},   
	pdfproducer={}, 
	pdfkeywords={keyword1} {key2} {key3}, 
	pdfnewwindow=true,      
	colorlinks=true,       
	linkcolor=OliveGreen, 
	citecolor=NavyBlue,        
	filecolor=magneta,      
	urlcolor=cyan           
}

\numberwithin{equation}{section}

\newcommand*\samethanks[1][\value{footnote}]{\footnotemark}

\newcommand\beal{\begin{equation}\begin{aligned}}
		\newcommand\eeal{\end{aligned}\end{equation}}

\begin{document}

\normalem

\setcounter{page}0
\def\ppnumber{\vbox{\baselineskip14pt
}}

\def\ppdate{
} 
\date{}

\title{\LARGE\bf Thermodynamics of dilute anyon gases \\ from fusion constraints}
\author{Yuto Nakajima$^1$, Umang Mehta$^{1,2}$, and Hart Goldman$^1$}
\affil{\it\small $^1$ School of Physics and Astronomy, University of Minnesota, Twin Cities, MN 55455, USA}
\affil{\it\small $^2$ Department of Physics and Center for Theory of Quantum Matter,\\
\it University of Colorado, Boulder, CO 80309, USA}
\maketitle\thispagestyle{firstpage}

\begin{abstract}
Recent measurements on 2$d$ materials tuning between fractional quantum anomalous Hall phases and a plethora of correlated electronic states call for a detailed understanding of the dynamics of anyons. Here we develop a general theory of the statistical mechanics of anyon gases at finite temperature, valid in regimes where the anyons are sufficiently dilute and can be treated as weakly interacting particles. We find that with a minimal set of universal braiding and fusion data, along with information about the hierarchy of anyon gaps, it is possible to construct a distribution function for \emph{any} dilute anyon gas, as well as derive thermodynamic observables. Our results are built on an anyon exclusion principle manifesting as a constraint on fusion outcomes of physical states. Our approach unifies and streamlines a range of results for itinerant anyon models, from solvable lattice Hamiltonians to large-$N$ field theories.
\end{abstract}

\pagebreak
\tableofcontents 
\pagebreak 


\section{Introduction}

The discovery of the fractional quantum anomalous Hall (FQAH) effect, first in twisted MoTe$_2$ ($t$MoTe$_2$)~\cite{cai2023signatures, zeng2023thermodynamic,park2023observation,xu2023observation}, and then in moir\'{e} rhombohedral graphene multilayers~\cite{lu2024fractional,lu2025extended,dong2025},  has opened a new frontier of phenomena where fractionalization emerges alongside more conventional electronic phases of matter, such as ordinary metals; charge orders and generalized Wigner crystals; and superconductivity.
Following the measurement of superconductivity in crystalline rhombohedral graphene without hBN alignment~\cite{han_signatures_2025}, Ref.~\cite{Li2025} recently reported superconductivity situated between FQAH and Hall metal phases in $t$MoTe$_2$, constituting the first simultaneous observation of superconductivity and the fractional quantum Hall effect in the phase diagram of a single material. One tantalizing possibility that has stimulated much recent theoretical work~\cite{Zhang2025,kim2025topologicalchiral,Shi2024doping,divic2024,Shi2025-nj,Pichler2025,Shi2025plateau,nosov2025,zhang2025holon,huang2025} is that the observed superconductivity arises from the dynamics of anyons from the proximate FQAH phase.

These rapid experimental developments suggest not only that anyon dynamics could underpin the intriguing electronic phases already under study, but that it may be possible to engineer new kinds of ``anyonic quantum phases of matter''  inaccessible to traditional quantum Hall settings in strong perpendicular magnetic field. However, because many-anyon systems inherently exhibit strong correlations and long-ranged entanglement, even their elementary thermodynamic and transport properties have resisted general understanding. Our goal in this work is to make progress in this direction.

We start with an old question: What is the distribution function for a suitably dilute gas of anyons at finite temperature? Following early pioneering work on the statistical mechanics of multi-anyon systems~\cite{Arovas1985}, Haldane proposed a modified exclusion principle interpolating between Bose and Fermi statistics by allowing filled states to reduce the Hilbert space dimension by fractional values~\cite{Haldane1991-ln}. This ``fractional exclusion principle'' bears the unique property that the probability of occupying a given state depends on the occupation of all others, in contrast to its more traditional Bose and Fermi counterparts. After Haldane's initial work, the idea of fractional exclusion principles was followed up on extensively~\cite{Ramanathan1992,Wu1994-dg,Nayak1994,Murthy:1994yew,Ha1994,Isakov:1994zz,deVeigy1994,Polychronakos1995-kd,Isakov:1996aj,Schoutens:1997xh,Murthy1999-rf,Ardonne2001-wx} (for recent reviews, see Refs.~\cite{Ouvry2009-ii,Xiong:2022mll}). In particular, fractional exclusion statistics can occur in general dimensions, with many major successes in the study of exactly solvable 1$d$ models such as the Calogero-Sutherland model~\cite{Murthy:1994yew,Ha1994,Isakov:1994zz}. 

Proposals for anyon gas distribution functions based on Haldane's fractional exclusion statistics~\cite{Ramanathan1992,Wu1994-dg,Nayak1994,Schoutens:1997xh,Ardonne2001-wx} focused on gases of a single species of abelian anyon and worked in approximations where each anyon was treated as an independent fractional statistics particle. They also lacked important microscopic information, such as whether the local particles the anyons emerged out of were bosons or fermions. As such, it has remained unclear how to extend these results to generic many-anyon systems in the thermodynamic limit, obtained by doping a topological order. Nevertheless, similar exclusion principles have recently been re-discovered in the quantum field theory literature on Chern-Simons theories coupled to matter in the 't Hooft large-$N$ limit~\cite{Geracie2015-ya,Minwalla:2020ysu,Minwalla2022-pg}, which are exactly solvable models of (non-abelian) anyon gases~\cite{Giombi2011-xo,Ofer2012-thermal}, as well as in the context of string net models deformed away from exact solvability~\cite{Vidal:2021isf,Ritz-Zwilling2024}. These results indicate that a unified exclusion principle for anyons should be possible.

Here we propose a framework for the statistical mechanics of sufficiently dilute anyon gases arising from \textit{any} bosonic or fermionic, abelian or non-abelian topological order, in which exclusion principles are re-expressed as constraints on anyon fusion. Our central result is a general formula for the distribution function of a dilute gas of $N\rightarrow\infty$ anyons, $a_1$, which fuses into $|\ell|$ daughter anyons, $\{a_m\}$, with quantum dimensions, $\{d_{a_m}\}$,
\begin{equation}\label{eq:distr_general_intro}
    n(\epsilon) = \frac{ \sum_{m=0}^{|\ell|} m\,d_{a_m}\,y^m }{ \sum_{m=0}^{|\ell|}d_{a_m}\, y^m }\,,\qquad y= e^{-\beta (\epsilon - \mu)}\,.
\end{equation}
For the salient example of a gas of charge-$e/k$ Laughlin quasiparticles ($k>0$ odd), this formula becomes:
\begin{align}
\label{eq:abelian_dist_intro}
n_{k}(\epsilon)=-\frac{y}{1-y}\left(k-(k+1)\frac{y^k-1}{y^{k+1}-1}\right)\,,
\end{align}
which reduces to the Fermi-Dirac and Bose-Einstein distributions for $k=1$ and $k\rightarrow\infty$ respectively. Similar distribution functions have been obtained in earlier 't Hooft limit studies of non-abelian Chern-Simons theories coupled to matter~\cite{Geracie2015-ya,Minwalla:2020ysu,Minwalla2022-pg}. The result in Eq.~\eqref{eq:abelian_dist_intro} should be understood as the proper extrapolation of these results to abelian topological orders with finite $k$, which physically corresponds to neglect of braiding effects in favor of anyon fusion.

\begin{figure}[t]
\centering
\begin{minipage}[b]{0.9\columnwidth}
    \centering
\includegraphics[width=1\columnwidth]{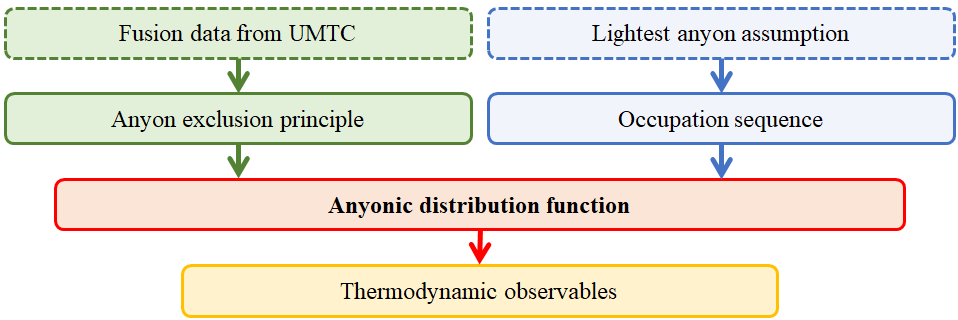} 
\end{minipage}
\caption{A schematic depiction of our statistical mechanical framework for dilute anyon gases. Universal fusion data leads to constraints on anyon occupancy of physical states, which we term an \emph{anyon exclusion principle}. In turn, an assumed ``lightest anyon'' allows us to construct an \emph{occupation sequence} of states to be summed over in the partition function. In the thermodynamic limit, these two ingredients are sufficient for deriving an anyon distribution function and thermodynamic observables for dilute anyon gases based on doping any input topological order.}
\label{fig:flowchart}
\end{figure}

Our result in Eq.~\eqref{eq:distr_general_intro} relies only on two sets of ingredients (see Figure~\ref{fig:flowchart}):
\begin{enumerate}
\item The anyon fusion data, which can be folded into a mathematical object called a \emph{unitary modular tensor category} (UMTC). In Eq.~\eqref{eq:distr_general_intro}, this data manifests in the quantum dimensions of the anyons, $a_m$, denoted $d_{a_m}$. 
\item The anyon with the smallest energy gap, which defines a prescription for constructing multi-anyon states. This prescription is encoded in a new object we term the \textit{occupation sequence}, $\ell=(a_0, a_1, \ldots,a_{|l|})$, of length $|\ell|+1$.
\end{enumerate}

\noindent
The first piece of data, the UMTC, is a universal classifying feature of the parent phase giving rise to the anyons~\cite{Barkeshli2019}. We show that the UMTC determines the exclusion properties of the anyon states in terms of the \textit{quantum dimensions} $d_a$ of the anyons, which are rational numbers that, loosely speaking, encode the degeneracy of an anyon type.

On the other hand, the second piece of data involves information about the anyon gaps, a property which is decidedly non-universal. However, we show that even just knowing the lightest anyon $a_1$ enables one to write down a family of  \emph{occupation sequences}, $\ell=(1,a_1,a_2,\ldots,a_{|\ell|})$, of possible multi-anyon states, in which the anyon $a_n$ determines the thermally allowed states of $n$ copies of the lightest anyon $a_1$, generalizing the requirement of multi-boson (-fermion) states being symmetric (antisymmetric). For abelian anyon gases, we find that the choice of occupation sequence is uniquely fixed by the statistics of the microscopic, local particles (electrons or bosons) the anyons emerge from. For non-abelian anyon gases an infinite family of occupation sequences is possible, each with quantitatively distinct thermodynamic observables.

With this data we can construct a partition function for the anyon gas and calculate thermodynamic quantities at intermediate temperature scales. The reason the lightest anyon is so essential is we expect it to determine whether the lightest local particle is a fermion (e.g. an electron) or a boson (e.g. a Cooper pair), meaning that it determines whether at low temperatures the system crosses over to metallic behavior characterized by the formation of a Fermi sea or to a Bose condensate. Recent work from other groups has also emphasized the important role played by the lightest anyon in determining the ground state properties of doped topological orders~\cite{Shi2024doping,Shi2025-nj,Shi2025plateau,nosov2025}.

Our work provides a general recipe (see Figure~\ref{fig:flowchart}) for constructing distribution functions for dilute anyon gases starting from a minimum set of categorical and dynamical information, allowing us to compute a range of thermodynamic observables such as entropy and pressure. Interestingly, because our approach emphasizes fusion constraints but treats braiding effects as sub-leading, our results depart from the early computation of the second virial coefficient for abelian anyon gases in Ref.~\cite{Arovas1985} while yielding new corrections to the virial expansion arising from fusion.

We proceed as follows. In Section~\ref{Sec2}, we begin with a discussion of how the dilute limit applies to anyons -- at least those which can be modeled as Chern-Simons gauge theories coupled to composite fermions or bosons --  arguing that at low charge densities there is a parametrically large temperature window where the partition function for an anyon gas can be modeled as one of nearly free particles with a constrained Hilbert space. In Section~\ref{Sec3}, we introduce our formalism, starting with an accessible introduction to UMTCs, and developing the notion of an occupation sequence. In Section~\ref{sec:distr_fn_occupation_sequence_sec4}, we leverage this framework to reach our general result for anyon distribution functions and discuss observable properties of some abelian anyon gases based on Laughlin states. Finally, in Section~\ref{sec:Examples_sec5}, we generate distribution functions for a range of abelian and non-abelian examples, finding anyonic analogues of Bose condensates and Fermi seas. We conclude in Section~\ref{Sec6}.


\section{Dilute limits of anyon gases}
\label{Sec2}

We begin by defining the concrete physical setting where our conclusions apply. Our focus is on situations in which anyon gases may be treated as \emph{dilute}, meaning that they reside at sufficiently low densities or high enough temperatures such that they may be treated as essentially free particles, but with modified exclusion statistics.

More precisely, we identify a regime where the inter-anyon interactions becomre short-ranged, and the anyon partition function can be approximated as a product of an  ``interacting'' part, containing information about local interactions, and an ``ideal'' part depending only on the anyons' dispersion, along with their fractional statistics and fusion properties. We may express the partition function as
\begin{align}
\label{eq:partition-fnc-factorization}
    Z=Z_{\mathrm{int}}\, Z_{\mathrm{ideal}}\,,\qquad Z_{\mathrm{ideal}}=\sum_{\{n_a\}}\sum_{\{\epsilon_a\}}\mathcal{N}(\{n_a\})\exp\left(-\beta \sum_{a} n_a(\epsilon_a-\mu) \right).
\end{align}
Here $n_a$ is the occupation number for an anyon $a$, the $\epsilon_a$'s are anyon energy levels depending on microscopic details of the model, and $\mathcal{N}(\{n_a\})$ are termed the fusion-singlet coefficients, which may be thought of as counting the number of configurations where the anyon species $\{a\}$ can occupy states with energies $\{\epsilon_a\}$, with corresponding occupation numbers $\{n_a\}$. They are the central objects implementing constraints on allowed multi-anyon states. We will define these objects in detail in Section~\ref{Sec3}. For the purposes of this Section, one can think of them as simply implementing a sum over physical states.

Even though anyons by their nature exhibit long-ranged braiding interactions, we expect there to generally be a temperature window where Eq.~\eqref{sec:partition-factorization} is a good approximation, with corrections due to braiding being sub-leading. We can motivate this dilute regime's existence by considering an explicit field theory model of an abelian anyon gas forming a superfluid (or superconductor) at low energies~\cite{Laughlin1988,Fetter1989,Lee1989,Chen1989,Fradkin1990a}. A more quantitative analysis of a broader class of abelian and non-abelian anyon gases is given in Appendix~\ref{app:diluteness}, reaching the same basic conclusions.

The theory we consider here\footnote{Note that the microscopic particles from which this anyon gas emerges may be either fermions or bosons, depending on whether $k$ is even or odd, on charge assignments, and on the couplings to the background gauge field. See Ref.~\cite{Shi2024doping} for a detailed discussion.} consists of fermions, $\psi$, coupled to an abelian Chern-Simons gauge field, $a_\mu$,
\begin{align}
\label{eq: abelian anyongas example}
\mathcal{L}=\mathcal{L}_\psi-\frac{k}{4\pi}ada+\mu\,\psi^\dagger\psi-\frac{1}{4g^2}f_{\mu\nu}^2+\dots\,,
\end{align}
where $\mathcal{L}_\psi$ is a non-relativistic Lagrangian for free fermions with effective mass $m_\star$ and we use the notation $ada=\varepsilon^{\mu\nu\lambda}a_\mu\partial_\nu a_\lambda$. This model describes a gas of anyons with exchange statistics ${\theta=\pi(1+1/k)}$. We define the anyon density as $\rho_\psi=\langle\psi^\dagger\psi\rangle$, and we do not include any background magnetic field.

At low temperatures, as in flux attachment, gauge fluctuations cause the fermions to each experience $+1/k$ units of magnetic flux, and they perfectly fill $k$ Landau levels with cyclotron gap proportional to the density, $\omega_c=\rho_\psi/(k\,m_\star)$. On integrating out the fermion Landau levels, a level $+k$ Chern-Simons term is generated that cancels the term native to the Lagrangian, leading to an effective superfluid action with Goldstone dual to $a_\mu$. 

As the system is heated up well beyond the cyclotron gap, 
\begin{align}
\frac{1}{k}\frac{\rho_\psi}{m_\star}\ll T\,,
\end{align}
fermions can be readily excited between Landau levels. The Landau level picture then ``melts'' and gives way to an anyon plasma~\cite{Lopez1993}. Here the electrostatic gauge fluctuations are Debye screened, precluding the cancellation of Chern-Simons terms that leads to superfluidity. More technically, the effective action after integrating out the $\psi$-fermions can be expressed as a derivative expansion controlled by the Debye mass of the Chern-Simons gauge field. This should allow the partition function to factorize as in Eq.~\eqref{eq:partition-fnc-factorization} via a saddle point approximation, with the ideal factor determined by the pure Chern-Simons theory and the corrections from interactions coming from higher derivative terms. See Appendix~\ref{app:diluteness} for a more detailed, complementary derivation of the diluteness condition in the context of $\mathrm{SU}(k)$ Chern-Simons theories coupled to fermions.

We therefore conclude that in the dilute regime the long-ranged interactions mediated by gauge fluctuations are suppressed, and the matter degrees of freedom can be thought of as nearly free particles of charge-$e/k$ with renormalized dispersion, $\epsilon(\bs{p})$. Furthermore, the suppression of braiding interactions can be applied explicitly through taking the large-$k$ limit, which renders the strength of the braiding interaction vanishingly small.

The remaining effect of the Chern-Simons gauge field is to constrain the physical Hilbert space to sum over gauge invariant states, which are Wilson loop configurations (the gauge invariant many-anyon states) in the corresponding Chern-Simons theory. These gauge invariant many-anyon states are indexed by the coefficients, $\mathcal{N}$. In this temperature regime, the partition function of the anyon gas can be approximated the ideal gas partition function in Eq.~\eqref{eq:partition-fnc-factorization}.

It is important to note that the dilute window does not extend to infinite temperature, although in Appendix~\ref{app:diluteness} we explain how in some special non-abelian cases it can extend down to $T=0$. An upper bound is set by the energy scale, $g^2$, of the Maxwell term, or by the overall gap to the topological order, whichever is smaller.

The existence of a temperature window where Eq.~\eqref{eq:partition-fnc-factorization} holds may come as a surprise, and it is natural to wonder just how useful it is. After all, systems with anyon excitations exhibit long-range entanglement, which we have not completely eliminated. Indeed, even if a partition function satisfies Eq.~\eqref{eq:partition-fnc-factorization}, the probability for an anyon to occupy a given state can still depend on the occupation of all the other states, and the partition function need not factorize into products of single-energy level partition functions, $z_i$, as ${Z_{\mathrm{ideal}}(\mu) = \prod_i z_i(\mu)}$, where the index $i$ labels the available energy levels. Nevertheless, we will see in the coming Section that the fusion coefficients do factorize in the the thermodynamic limit, enabling one to rewrite $Z_{\mathrm{ideal}}$ in terms of one-particle partition functions depending on fusion data alone.


\section{Fusion constraints and anyonic exclusion principles}
\label{Sec3}

We have argued that heating up a many-anyon system allows us to approximate its partition function as a product of free and interacting parts, $Z_{\mathrm{int}}Z_{\mathrm{ideal}}$, as we might expect for a dilute gas of weakly interacting particles. But while for bosons or fermions the ``free'' factor, $Z_{\mathrm{ideal}}$, is straightforward to write down by leveraging the Pauli exclusion principle or lack thereof, the analogous occupancy constraints on anyons are less obvious, since anyon species can fuse into one another. We therefore proceed to develop a general statistical mechanical theory allowing us to compute $Z_{\mathrm{ideal}}$ for \emph{any} dilute anyon gas. Along the way, we will introduce new formalism based on the now mature understanding of gapped topological orders in terms of tensor categories (for a comprehensive review, see Ref.~\cite{Barkeshli2019}). Surprisingly, we will see that this mathematical structure can be leveraged not only to classify ground state properties of topological orders but also to study their thermodynamic properties under finite temperature and doping.


\subsection{Categorical preliminaries}

We begin with a physically grounded review of the mathematical framework of \textit{tensor categories} for organizing anyon data. For further details, see Appendix~\ref{app:UMTC}.


\subsubsection{Unitary Modular Tensor Categories (UMTCs)}

The basic properties of anyons are their (abelian or non-abelian) braiding statistics -- how the wave function transforms when one anyon is exchanged with another -- and their fusion outcomes -- two anyons may be brought together to form a third. These two sets of data furnish a \textit{unitary modular tensor category (UMTC)}.
UMTCs are mathematical objects consisting of (1) a finite set, $\mathcal{C}$, of anyons labeled by $a,b,c,\ldots \in \mathcal{C}$, (2) the fusion coefficients $N^c_{ab}$, (3) the $F$-symbols $[F^{abc}_d]_{ef}$, and (4) the $R$-symbols $R^{ab}_c$. Throughout this work, Latin indices at the beginning of the alphabet, $a,b,c,\dots$, will be used to denote anyon species.
 
In traditional circumstances, anyons are understood to be static topological line operators -- or defects -- that can be inserted into a model of a gapped, topologically ordered phase. Anyons can be ``multiplied together,'' in the sense that two anyon lines can be brought together and fused into a third. The results of this multiplication define the fusion coefficients,
 \begin{equation}
    \label{eq:DefFusion}
     a \times b = \sum_{c \in \mathcal{C}} N^c_{ab} \ c,
 \end{equation}
where the $N^c_{ab}$'s are non-negative integers counting the number of anyons of species $c$ formed from fusing $a$ with $b$. To the fusion $a \times b \to c$, the vector space $V_{ab}^c$ is assigned, whose dimension is $\dim V_{ab}^c = N_{ab}^c$. A set of anyons is \textit{abelian} is the fusion of any two of them gives only a single result in the sum on the right hand side of Eq.~\eqref{eq:DefFusion}.

Because every system exhibiting fractionalization at long wavelengths emerged from a microscopic system of local bosons or fermions, one must introduce the trivial anyon, denoted $1$, which is a truly local particle whose insertion is inconsequential to the ground state topological order. Fusion of the trivial anyon with any other is then simply given by ${1 \times a = a \times 1=a}$ for an arbitrary $a$. The inverse $\overline{a}$ for $a$ is defined as the unique object such that
\begin{equation}
    a\times \overline{a} = 1 + \ldots = \overline{a}\times a\,,
\end{equation}
meaning $N_{a\overline{a}}^1 = N_{\overline{a}a}^1 = 1$.

These fusion rules imply that the anyons that form the UMTC need not only be interpreted as single particles, but also as particles existing in the background of a number of local bosons. The trivial anyon 1 is thus also called the \textit{transparent boson}. It can represent the vacuum with no anyons or a state with multiple copies of the local anyon.

Any anyon can be formed into a loop, which can be associated with a \textit{quantum dimension}, $d_a>0$. Conveniently, like the anyons themselves, the quantum dimensions also obey the fusion equation
\begin{equation}
    d_a d_b = \sum_c N_{ab}^c d_c\,.
\end{equation}
One can therefore observe that the trivial anyon evidently has $d_1 = 1$ and the quantum dimension of an anyon is the same as that of its inverse, $d_a = d_{\overline{a}}$. The topological order also has its own \textit{total quantum dimension} $\mathcal{D}>0$ given by
\begin{equation}
    \mathcal{D}^2 = \sum_a d_a^2\,.
\end{equation}
Note that quantum dimensions are \textit{not} required to be integers. 

The quantum dimensions, $d_a$, admit a simple interpretation as the effective ``degeneracy'' of an anyon species. This analogy can be made especially precise in examples where anyons are ``deformations'' of quantum spins, such as $\mathrm{SU}(2)_k$ Read-Rezayi states. Start with a single spin with angular momentum $j$. This spin lives in an irreducible representation of $\mathrm{SU}(2)$ and consists of $2j+1$ components, characterizing the degeneracy of a spin $j$ state. To transmute this spin into an anyon, we deform the $\mathrm{SU}(2)$ spin to one in $\mathrm{SU}(2)_k$. The anyons in $\mathrm{SU}(2)_k$ are still labeled by a half-integer $j$, but now with the added restriction that $j\le k/2$. Correspondingly, the quantum dimensions of these anyons asymptotically approach the degeneracy of the spin-$j$ representation in the $k\rightarrow\infty$ limit~\cite{bonderson2007non},
\begin{equation}
    d^{SU(2)_k}_j = \frac{ \sin \frac{(2j+1)\pi}{k+2} }{ \sin \frac{\pi}{k+2} } \xrightarrow{k\rightarrow\infty} 2j+1\,,
\end{equation}
from which we can conclude that the quantum dimensions, loosely speaking, are a measure of the degeneracy of the respective anyons.

Rounding out the data associated with a UMTC -- thereby uniquely characterizing the topological order -- are the braiding of anyons with one another along with the associativity of fusion. These are respectively encoded in the $R$ and $F$-symbols. However, in the dilute regime, we find that the fusion coefficients suffice for constructing the partition function\footnote{Strictly speaking, we find that our results only require the more general concept of a \emph{commutative fusion category}.}. We will comment on the physical interpretation of this result later.


\subsubsection{Fusion-singlet coefficients}

Because fusion coefficients and quantum dimensions contain the information about the degeneracies of anyon states, it is natural to consider the following counting problem. If one starts with a set of $n$ anyons, $a_1,\dots,a_n$, how ways can local particles, i.e. trivial anyons, be created by fusing all of them? On a 2-dimensional spatial surface of genus $g$, the result is given by a special fusion coefficient, the \emph{singlet coefficient}, $\mathcal{N}_g(a_1,\dots,a_n)$, which, in the limit of large\footnote{More precisely, this formula holds when $n\gg \sum_a M_a$ where $M_a$ is the smallest positive integer such that $a^{M_a}$ contains $1$. For $n>\sum_a M_a$, the fusion of all the anyons is guaranteed to have at least one trivial anyon.} $n$ satisfies the asymptotic formula~\cite{Pasquier1987,Verlinde:1988sn},
\begin{equation}
\label{eq:VerlindeAsymptotic}
    \mathcal{N}_g \left(a_1,\ldots,a_n\right) \sim \mathcal{D}^{2g-2} \prod_{i=1}^n d_{a_i}.
\end{equation}
This formula has a natural interpretation on the torus, where the normalizing factor is unity, as it is simply the product of the degeneracies of each participating anyon species.

This fusion coefficient is among the most important quantities for our formalism, as well as for calculating ground state degeneracy of the topological order, and we will dedicate Section \ref{sec:singlet_cond} to elucidating its physics.


\subsubsection{Incorporating local fermions}

We have thus far eluded the fact that UMTCs only classify topological orders arising from systems of microscopic bosons, making the $1$-anyon a local boson. Nevertheless, for our purposes it is possible to borrow the basic structure of UMTCs to describe topological orders with local fermions. 

Starting with a bosonic topological order, we introduce a unique fermion, $\psi$, that braids trivially with every other anyon but fuses to the identity,
\begin{equation}
    \psi \times \psi = 1\,.
\end{equation}
Under this definition, $\psi$ may be regarded as a local particle with similar status to the identity anyon, $1$, which remains bosonic and may be understood as a ``Cooper pair'' of the electrons corresponding to $\psi$.

We now revisit the counting problem of the previous subsection. In addition to forming a local boson, we wish to allow our set of anyons $a_1,\dots,a_n$ to also form the local fermion, $\psi$. The singlet coefficient therefore becomes the sum of two corresponding fusion coefficients,
\begin{equation}\label{eq:fermion_singlet_coeff}
    \mathcal{N}_g (a_1,\ldots,a_n) = N^1_{a_1\ldots a_n} + N^\psi_{a_1\ldots a_n}\,,
\end{equation}
where $N^\psi$ ($N^1$) is the number of local fermions (bosons) formed through the fusion process. 

Although the introduction of a local fermion alters the exact form of $\mathcal{N}$ at small anyon number, in Appendix~\ref{app:UMTC}, we show explicitly that this sum also obeys the same asymptotic form as in  Eq.~\eqref{eq:VerlindeAsymptotic}. To arrive at this conclusion intuitively, we note that here we have constructed a non-trivial fermionic topological order by stacking a ``transparent'' fermion theory $\{1,\psi\}$ onto a UMTC, creating new anyons from old ones by composing them with the local fermion. The resulting fusion coefficients are identical to those of the underlying UMTC, provided one keeps track of fermion parity to replace $1$ by $\psi$ as needed. Explicitly, if $a_\psi$ is the composite of anyon $a$ in the original UMTC with the fermion, one should require
\begin{equation}
    N^c_{ab} = N^c_{a_\psi b_\psi} = N^{c_\psi}_{a_\psi b} = N^{c_\psi}_{a b_\psi}\,.
\end{equation}

We remark that some fermionic phases, such as the Pfaffian, cannot be obtained by stacking a transparent fermion onto a UMTC. However, these examples can always be quotiented by the transparent fermion theory, $\{ 1, \psi \}$,  to enable one to calculate the singlet coefficient in Eq.~\eqref{eq:fermion_singlet_coeff} from the vantage of a smaller bosonic topological order. Formally, the resulting quotient is not a UMTC. However, the asymptotic formula, Eq.~\eqref{eq:VerlindeAsymptotic}, can be shown even to hold for such theories.


\subsection{Anyonic exclusion as a singlet condition}\label{sec:singlet_cond}

We are now equipped to return to the problem of dilute anyon gases at finite temperature and density. Although the UMTC formalism of the previous section was developed to study gapped phases at zero temperature, it can be leveraged to establish occupancy rules for many-anyon dynamics.

\begin{figure}
\centering
\begin{minipage}[b]{0.48\columnwidth}
    \centering
    \includegraphics[height=0.8 \linewidth]{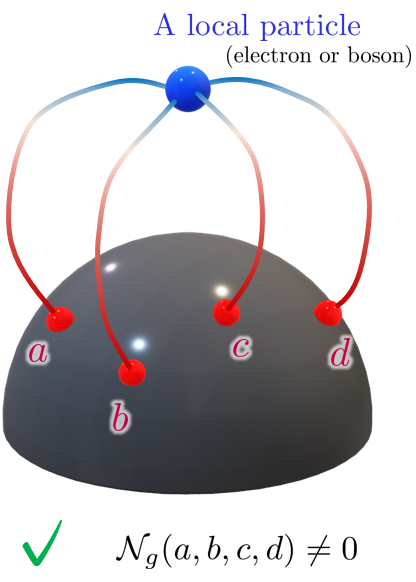} 
    \subcaption{Example of a physical state}
\end{minipage}
\begin{minipage}[b]{0.48\columnwidth}
    \centering
    \includegraphics[height=0.8 \linewidth]{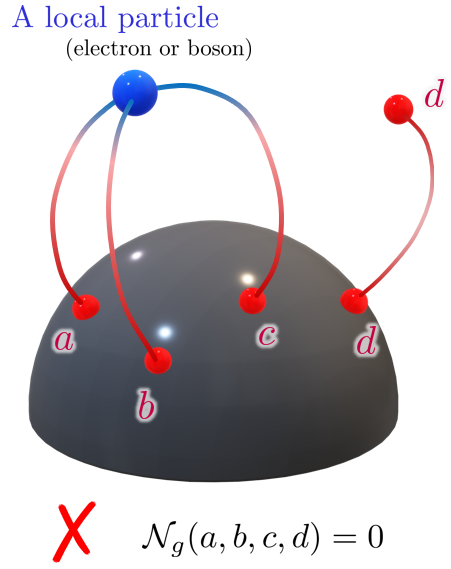}  
    \subcaption{Example of an unphysical state}
\end{minipage}
\caption{An example of \textbf{(a)} a physical state, where the anyons can fuse into a local particle, and \textbf{(b)} an unphysical state, where the anyons cannot.
The given state involving anyons $a$, $b$, $c$, and $d$ is allowed if and only if there exists a fusion channel into a local particle, i.e. $\mathcal{N}_g(a,b,c,d) \neq 0$.}\label{fig:Identity}
\end{figure}

Because anyons are fractionally charged particles arising from an underlying systems of microscopic electrons or bosons, they cannot exist in isolation. For example, inserting a single Laughlin quasiparticle into a $\nu=1/3$ fractional quantum Hall (FQH) state breaks gauge invariance and is thereby physically impossible. However, on inserting a single electron into the FQH state, that electron will delocalize into three anyons (energetics permitting). More generally, the relevant states for even many-anyon systems are the \emph{fusion singlets} alone. Heating the system while remaining below the FQH gap will not alter this conclusion. 

We recast this observation as a general selection rule for physical many-anyon states:\\

\noindent \textbf{Fusion singlet condition.} \textit{ A collection of anyons $a_1,\ldots,a_n$ can form a physical state if and only if the fusion of all of them consists of a trivial boson or fermion. In other words, a physical state consisting of anyons $a_1,\ldots,a_n$ must have $\mathcal{N}_g(a_1,\ldots,a_n) \ne 0$.}\\

\noindent
Recall that the singlet coefficient, $\mathcal{N}_g(a_1,\ldots,a_n)$, counts the number of ways  anyons $a_1,\ldots,a_n$ can fuse together to form local electrons or bosons. Conversely, it is the number of ways a collection of local particles can split into a deconfined collection of anyons $a_1,\ldots,a_n$. 
Examples of the allowed and disallowed states are shown in Figure~\ref{fig:Identity}. That occupancy constraints for anyons could be expressed in terms of fusion coefficients was recognized relatively recently in the study of Chern-Simons-matter theories~\cite{Minwalla2022-pg} and string-net models~\cite{Vidal:2021isf,Ritz-Zwilling2024}, but we emphasize that it applies generally beyond the realm of solvable models and is mandated by gauge invariance. 

Let us unpack this condition for an anyon gas. Imagine injecting the ground state of a topologically ordered system with local particles. The doped particles will split into anyon excitations in a manner compatible with the singlet condition, with the most stable configurations depending on the hierarchy of anyon gaps. Any physical state that consists of a collection of anyons must be obtained by such a procedure, which, if done adiabatically, can be reversed to fuse those anyons back into a collection of local particles. Hence, the state consisting of a collection of anyons is physically realizeable if and only if $\mathcal{N}_g \ne 0$.

As an anyon occupation constraint, the fusion singlet condition constitutes a kind of \emph{anyonic exclusion principle}, and it is our starting point for a statistical mechanics of dilute anyon gases. Note that this anyonic exclusion principle is distinct from earlier notions of fractional exclusion statistics -- we comment on this distinction in more detail in Section~\ref{sec:distr_fn_occupation_sequence_sec4}.

Taken alone, our anyonic exclusion principle lacks one key piece of information: the exclusion statistics of the microscopic, local particles the anyons fuse to. We now turn to the task of ameliorating this issue, which requires some additionally requires some basic assumptions about the anyon spectrum.


\subsection{Introducing dynamics: Occupation sequence formalism}
\label{Sec:Sequence}

\subsubsection{The lightest anyon}

The analysis thus far has focused on kinematic constraints on the anyon Hilbert space. But to construct a thermal partition function, it is necessary to make some basic assumptions about the anyon energy spectrum and dynamics. 

Anyon gases arise on doping topological orders with physical charges -- local electrons or bosons -- which in turn split into anyons. The relative concentrations of different anyon species are determined by the hierarchy of anyon gaps. For now, we will assume that there is a unique anyon, $a$, with the smallest gap in the theory, $\Delta_\mathrm{0}$. We will imagine that the system is doped above $\Delta_0$ with a chemical potential, $\mu$, but at temperatures still much smaller than the gap of the next-lightest anyon:
\begin{equation}\label{eq:temp_for_single_flavour_gas_sec3}
    \Delta_\mathrm{0} < \mu < \Delta_\text{next-lightest}\,,\qquad T\ll \Delta_\text{next-lightest}\,.
\end{equation}
In this regime\footnote{We will always assume that $\Delta_\mathrm{0}$ is smaller than the overall gap to topological order, $\Delta_{\mathrm{TO}}$. However, there is no principle requiring this: Some topological orders may have $\Delta_\mathrm{TO}<\Delta_{0}$, in which case the topological order will be destroyed without nucleation of anyons. This distinction is analogous to the difference between Type I and Type II superconductors, where here anyons play the role of vortices.}, the system only excites the lightest anyon and we obtain a gas of $a$ anyons, while energetically suppressing the other anyon types. We emphasize that our requirement that such a temperature window exists is for ease of presentation. We will describe in Section~\ref{sec:distr_fn_occupation_sequence_sec4} how our formalism can be extended to mixtures of different anyon types.

In many situations, the lightest anyon will also determine the charge of the lightest local particle, assuming that the binding energy of the corresponding local particle is comparable in magnitude to the anyon gap (although this need not be a strict assumption). We now proceed to codify this statement by showing that the exclusion statistics of the lightest local particle plays a key role in determining whether the anyon gas ultimately behaves as a Fermi gas or Bose condensate as the temperature is decreased.

\subsubsection{The occupation sequence}

We now consider the problem of filling states with the lightest anyon. Knowing the fusion singlet condition, we must require that any physical state accommodate fusion of every anyon back to local particles. While this condition furnishes a fractional exclusion principle for the anyons, it is agnostic to the exclusion statistics of the local particles themselves. We therefore introduce the notion of an \emph{occupation sequence}, which encodes both the fractional exclusion statistics of the lightest anyon, along with the Bose or Fermi statistics of the local particles these anyons can form. The occupation sequence marries both universal topological data with dynamical information about the anyon spectrum and is among the central contributions of this work. Provided both the fusion singlet condition and an occupation sequence, one can construct a distribution function for \emph{any} dilute anyon gas.

The recipe for constructing an occupation sequence is as follows. We initiate the sequence with $1$ -- representing the empty state -- followed by the lightest anyon, denoted by $a_1$. We interpret the fusion of $a_1$ with itself as a two-particle state given by
\begin{equation}
    a_1 \times a_1 = \sum_b N^b_{1 1}\, b\,,
\end{equation}
where $b$ runs over the available anyons in the UMTC and we have adopted the notation (useful later) $N^{b}_{a_m a_n}\equiv N^b_{mn}$. We now choose one such anyon, $a_2$, so that up to a fusion coefficient $a_1\times a_1 = a_2 + \ldots$. If $a_1$ were a spin-1/2 particle, the allowed $b$'s in the above equation would correspond to a spin-singlet (antisymmetric) or spin-triplet (symmetric) state, so a choice of $b$ is indeed the anyonic generalization of two-particle occupation statistics.

We next construct a three-particle state, formally represented by the fusion,
\begin{equation}
    a_1 \times a_2 = \sum_c N^c_{12}\, c\,,
\end{equation}
and pick one such $c$ to be $a_3$. For $a_4$, representing the four particle state, there are two possible fusion channels,
\begin{equation}
    \begin{split}
        a_1\times a_3 &= \sum_e N^e_{13}\, e\,,\\
        a_2\times a_2 &= \sum_g N^g_{2 2}\, g\,.
    \end{split}
\end{equation}
To assign $a_4$, we pick an anyon that is common to both fusions above. Associativity of fusion guarantees that at least one such anyon exists. Under this prescription, the sequence can in principle be continued indefinitely by picking new anyons such that 
\begin{equation}
    a_n \times a_m = N^{m+n}_{mn} a_{n+m} + \ldots
\end{equation}
for all $n$ and $m$. 

\begin{figure}
    \centering
    \includegraphics[width=0.5 \linewidth]{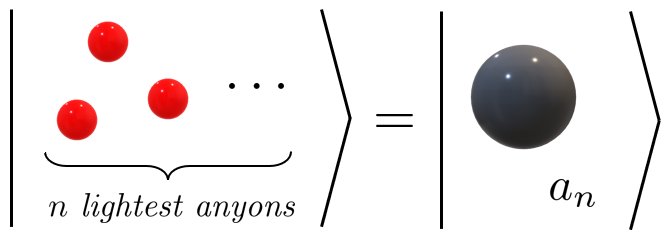}
    \caption{The fusion channel $a_n$ defines the state in which $n$ lightest anyons occupy a given energy level.}
    \label{fig:OccupyAnyon}
\end{figure}

Iterating this process leads to an infinite \emph{occupation sequence} (see Figure~\ref{fig:OccupyAnyon}), 
\begin{align}
\label{eq:occupation_sequence}
    \ell = (1, a_1, a_2, \ldots), \qquad a_n\times a_m = N^{m+n}_{mn} a_{n+m} + \ldots
\end{align}
However, if at any point along the sequence we encounter a local particle, $a_{\mathrm{local}}$, be it fermionic or bosonic, we can \textit{choose} to end the sequence there,
\begin{equation}\label{eq:occupation_sequence_truncated}
    \ell = (1, a_1, a_2, \ldots , a_{\mathrm{local}})\,.
\end{equation}
This sequence may be interpreted as starting with an empty state and proceeding to fill it with quanta of the lightest anyon until they are capable of fusing to a single local boson or fermion. The truncation at a local particle then has the following physical meaning: If $a_{\mathrm{local}}$ is a fermion, the truncation simply implements Pauli exclusion and is thus mandated on fundamental grounds. 

On the other hand, if $a_{\mathrm{local}}$ is a boson, prohibiting further occupation of a state must be implemented energetically through interactions and amounts to a ``hard-core'' constraint\footnote{Note that one cannot \emph{a priori} exclude sequences truncating at anyons. These sequences would correspond to situations where strong interaction effects prefer partial occupation of states in a manner consistent with the fusion singlet condition. These situations are beyond the scope of this work, but an example is discussed in Appendix~\ref{app:tHooft}.} For bosons, then, one could also imagine repeating the sequence indefinitely: The next appearance of $a_\mathrm{local}$ represents a state occupied by two copies of the local anyon (fractionalized into $a_1$'s), and so on for all further entries.

We remark that the constraints on multi-anyon states encoded by the occupation sequence can be associated with both kinematic and dynamic properties of the microscopic theory. For example, if the microscopic realization of the anyon theory is a Chern-Simons gauge theory coupled to matter, the specific realization of matter -- fermions or bosons -- determines the fusion channel of the allowed multi-anyon states, and the occupation sequence encodes this additional data. Additionally, different fusion channels can have different energetic costs, and one generally expects the lightest fusion channel to be preferred. Indeed, for non-abelian anyon gases, we will find that large families of occupation sequences are possible and lead to quantitatively distinct physical properties. In these cases, our construction is best viewed as a classification framework for the different types of physically realizable anyon gas partition functions.


We have now established the two ingredients required to describe anyon statistical mechanics: the fusion singlet condition, which provides a means of counting valid multi-anyon states, and the occupation sequence, which generalizes wavefunction symmetry and antisymmetry of bosons and fermions. With these tools, we now turn our attention back to the partition function for the dilute anyon gas.

\section{Distribution functions for dilute anyon gases}\label{sec:distr_fn_occupation_sequence_sec4}


\subsection{Diluteness}

In Section~\ref{Sec2}, we motivated the existence of a parametrically large temperature window in which anyon gases are \emph{dilute}, in the sense that they behave as weakly interacting gases of dispersing particles with anyonic exclusion statistics determined by fusion constraints. In this regime, the partition function factorizes into what we term ideal and interacting parts, 
\begin{equation}
\label{eq: diluteness, again}
    Z \approx Z_{\mathrm{ideal}}[\text{anyon data, spectrum}] \times Z_\mathrm{int}[\text{couplings}]\,.
\end{equation}
The ideal part, $Z_{\mathrm{ideal}}$, describes anyons' occupancy of (possibly renormalized) single-particle Fock states, while the interacting part $Z_{\mathrm{int}}$ depends on the detailed short-ranged interactions. Although specific anyon gases will vary widely in their low temperature behavior and hence in the specific temperature window where they become dilute, Eq.~\eqref{eq: diluteness, again} is a general definition applying to any system. 

The ideal partition function describes a \textit{minimally interacting anyon gas}, in which the statistical interaction is encoded by anyon data and doesn't further renormalize the the anyons' single-particle energies. It is the object of primary interest to us, and we will spend the rest of this work computing it in many specific examples. We leave understanding corrections due to $Z_{\mathrm{int}}$ to future work.


\subsection{The ideal partition function}\label{sec:partition-factorization}

Now consider a general dilute anyon gas with lightest anyon, $a_1$, whose (possibly degenerate) energy spectrum is $\{\epsilon_i\}$, where $i$ runs over all symmetry quantum numbers.

In a given configuration of $a_1$'s, we define the occupation number, $n_i\geq0$, to count the number of anyons in a state with energy $\epsilon_i$. According to the fusion singlet condition, such a configuration is only physical if the anyons occupying each state can be fused back into local, microscopic particles. In addition, the anyons in each state furnish an occupation sequence, $\ell$, which places an upper bound, $n_i\leq|\ell|$, if the sequence is finite (i.e. if the local particle truncating the sequence is a fermion or a ``hard-core'' boson). Here we have introduced the notation, $|\ell|$, to denote the index of the last anyon in the occupation sequence. In other words, the total cardinality of the occupation sequence is $1+|\ell|$.

The above considerations immediately imply an expression for the partition function,
\begin{equation}\label{eq:kin_part_fn_defn_sec3}
    Z_{\mathrm{ideal}} = \sum_{n_i=0}^{|\ell|} \mathcal{N}_g (a_{n_1}, a_{n_2}, \ldots) \exp \left( -\beta \sum_i n_i (\epsilon_i - \mu) \right)\,,
\end{equation}
where $\mathcal{N}_g$ counts the number of available configurations for the anyons in fusion channels, $\{ a_{n_i} \}$, to occupy. For all possible \emph{abelian} occupation sequences, one can further make the simplifying replacement,
\begin{equation}\label{eq:kin_part_fn_unconstrained_sec3}
    Z_{\mathrm{ideal}} = \sum_{n_i = 0}^{|\ell|} \mathcal{N}_g(\, \underbrace{a_1,\ldots,a_1}_{\sum_i n_i\text{ times}}\,) \exp \left( -\beta \sum_i n_i (\epsilon_i - \mu) \right)\,.
\end{equation}
However, for \emph{non-abelian} occupation sequences we emphasize that
\begin{equation}
    \mathcal{N}_g(a_{n_1},a_{n_2},\ldots) < \mathcal{N}_g(\, \underbrace{a_1,\ldots,a_1}_{\sum_i n_i\text{ times}}\,)\,,
\end{equation}
exemplifying how the occupation sequence restricts the number of allowed multi-anyon states beyond just the singlet condition.

To clarify the meaning of the occupation sequences, the partition function in Eq.~\eqref{eq:kin_part_fn_defn_sec3} can be seen to capture free, spinless gases of ordinary bosons or fermions. The occupation sequences of free Bose and Fermi gases are respectively given by
\begin{equation}\label{eq:Bose_Fermi_occ_sequence}
    \begin{split}
        \ell_\mathrm{Bose} &= ( 1,1,1,\ldots ),\\
        \ell_\mathrm{Fermi} &= ( 1, \psi )\,,
    \end{split}
\end{equation}
where $\psi$ is a local fermion. In both cases, the fusion-singlet coefficients are always $1$. The infinite length of the Bose sequence allows occupation numbers to be unbounded, while the finiteness of the Fermi sequence enforces Pauli exclusion by requiring $n_i = 0$ or $1$.

These cases exemplify how the length of an occupation sequence determines the behavior of an anyon gas as one extrapolates to  low temperatures. Infinitely long occupation sequences, which are allowed if the lightest local particle is a boson, will exhibit Bose condensates by enabling infinitely many anyons to occupy a single state (compatibly with the selection rules). On the other hand, sequences encountering a local fermion will truncate the occupation number, leading to an anyonic generalization of a Fermi sea. 

The partition function in equation \eqref{eq:kin_part_fn_defn_sec3} can be simplified further in the limit of large number of anyons, $\langle N \rangle = T \frac{\partial}{\partial\mu} \log Z_{\mathrm{ideal}} \gg 1$. In this limit, the sum over occupation numbers is dominated by configurations with ${\sum_i {n_i} \gg 1}$, allowing us to make use of the asymptotic formula from Eq.~\eqref{eq:VerlindeAsymptotic},
\begin{equation}
    \mathcal{N}_g (a_{n_1},a_{n_2},\ldots) \sim \mathcal{D}^{2g-2} \prod_i d_{a_{n_i}}\,.
\end{equation}
Using this formula, we can further factorize the ideal partition function into partition functions associated with each individual state, $z(\epsilon_i)$,
\begin{equation}\label{eq:single_level_part_fn}
    \begin{split}
        Z_{\mathrm{ideal}} &\approx \mathcal{D}^{2g-2} \prod_i z(\epsilon_i)\,,\\
        z(\epsilon_i) &=  \sum_{n=0}^{|\ell|} d_{a_n} e^{-\beta n(\epsilon_i - \mu)} \,,
    \end{split}
\end{equation}
demonstrating that, in the thermodynamic limit, dilute gases anyons behave as essentially free particles with ``degeneracy'' given by their quantum dimensions, each occupying states with capacity $|\ell|$. 


\subsection{Anyonic distribution functions}\label{sec:gen_distr_fn_sec4.2}

The single-state partition function, $z(\epsilon)$, from Eq.~\eqref{eq:single_level_part_fn} can be leveraged to calculate the average occupation number of a state with energy $\epsilon$ using the standard relation,
\begin{equation}
    n(\epsilon) = \frac{1}{\beta} \frac{\partial}{\partial\mu} \log z(\epsilon)\,,
\end{equation}
Defining the Boltzmann weight, $y\equiv e^{-\beta(\epsilon-\mu)}$, we thereby obtain a formula for the general distribution function of any dilute anyon gas, 
\begin{equation}\label{eq:distr_general_sec4}
    n_\ell(\epsilon) = \frac{ \sum_{m=0}^{|\ell|} m\, d_{a_m} y^m }{ \sum_{m=0}^{|\ell|} d_{a_m} y^m }\,,
\end{equation}
which depends only on the fusion data, in the form of the quantum dimensions, and the hierarchy of anyon gaps, in the appropriate choice of occupation sequence. This general formula for the distribution function of a dilute anyon gas is the central result of this work.

The anyon distribution function \eqref{eq:distr_general_sec4} has several important implications for anyon thermodynamics, many of which are apparent even for gases of abelian anyons, our results for which we introduce here. Because the quantum dimensions of abelian anyons are unity, the key data distinguishing them from ordinary Bose or Fermi gases lies in the occupation sequence. Serendipitously, the occupation sequence for an abelian anyon gas is uniquely determined by the lightest anyon, due to the absence of branching fusion channels.

For an abelian anyon gas, if the local particles on the occupation sequence are all bosons, then the occupation sequence can be chosen to be infinite, exactly reproducing the Bose-Einstein distribution,
\begin{equation}
    n^\mathrm{abelian}_{\ell\rightarrow\infty}(\epsilon) = n_\mathrm{Bose}(\epsilon) = \frac{y}{1 - y}\,.
\end{equation}
Hence, this abelian anyon gas is therefore thermodynamically indistinguishable from a Bose-Einstein condensate at all temperatures below the gap to the next lightest anyon. This result is intuitive: If the lightest anyons fuse to local bosons, the natural expectation would be for the bosons to condense at low temperatures. Such an outcome applies not only to bosonic topological orders, but also to fermionic topological orders whose lightest local particle is actually a bound state of fermions with even charge (such as a Cooper pair).

\begin{figure}[t]
\centering
\begin{minipage}[b]{0.48\columnwidth}
    \centering
    \includegraphics[width=1\columnwidth]{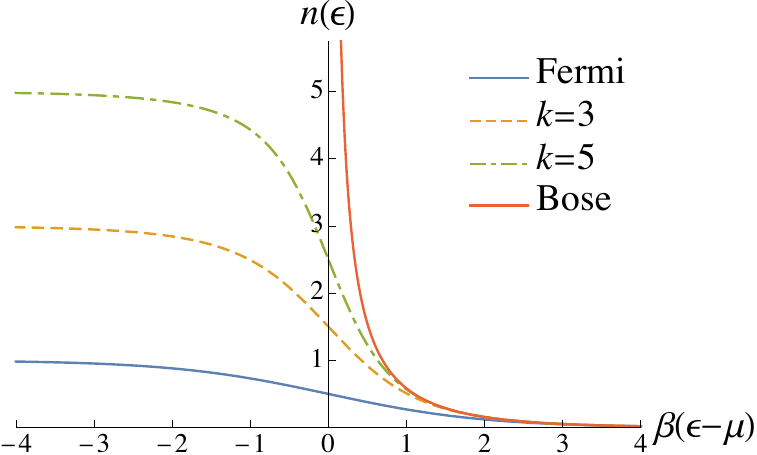}
    \subcaption{Raw distribition functions}\label{fig:DistNonScaledAbelian}
\end{minipage}
\begin{minipage}[b]{0.48\columnwidth}
    \centering
    \includegraphics[width=1\columnwidth]{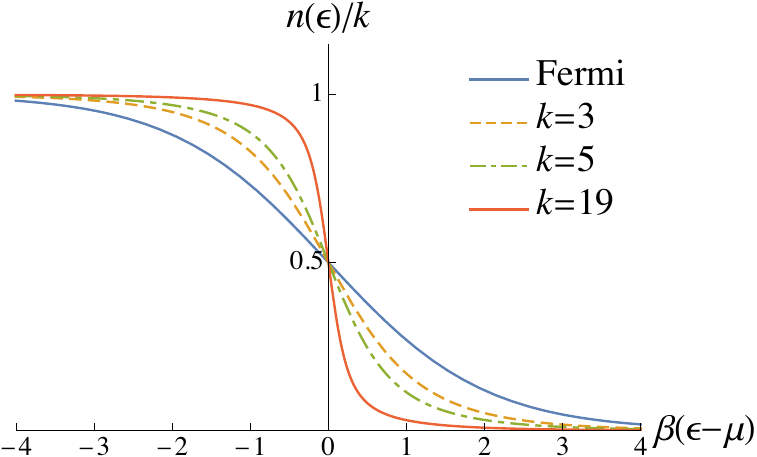}  
    \subcaption{Distribution functions scaled by $k$}\label{fig:DistScaledAbelian}
\end{minipage}
\caption{The distribution functions, Eq.~\eqref{eq:abelian_finite_distr_fn_sec4}, for various abelian anyon gases of charges $e/k$, featured alongside the Bose-Einstein ($k\rightarrow\infty$) and Fermi-Dirac ($k=1$) distributions.
\textbf{(a)} Each distribution function reaches a maximum at $k$, the length of the occupation sequence. \textbf{(b)} The distribution functions rescaled by $k$, showing the occupancy in units of local particles. The shape of the distribution depends sensitively on $k$.}
\label{fig:distfuncs_Sec4}
\end{figure}

If the first local particle to appear on the occupation sequence is a fermion, the sequence is truncated to enforce Pauli exclusion. In these cases, the distribution function only depends on the length of the sequence, $|\ell|\geq 1$, corresponding to the minimum number of lightest anyons required to form a local particle. Let us (suggestively) relabel this number by ${|\ell|\equiv k}$. The distribution function of the resulting abelian anyon gas resembles the Fermi-Dirac distribution but with quantitative differences,
\begin{equation}\label{eq:abelian_finite_distr_fn_sec4}
    \begin{split}
        n^\mathrm{abelian}_k(\epsilon) &= \frac{\sum_{m=0}^k m\,y^m}{\sum_{m=0}^k y^m}\\
        &= -\frac{y}{1 - y} \left( k - (k+1) \frac{ y^k - 1 }{ y^{k+1} - 1 } \right)\,,
    \end{split}
\end{equation}
where we continue to use the notation for the Boltzmann weight, ${y\equiv e^{-\beta(\epsilon-\mu)}}$. For odd $k$, this abelian distribution function corresponds to a gas of $e/k$ anyons doped into a ${\nu=1/k}$ Laughlin state of electrons. The distribution function in Eq.~\eqref{eq:abelian_finite_distr_fn_sec4} interpolates between Fermi-Dirac and Bose-Einstein distributions, which are respectively recovered for $k=1$ and $k\rightarrow\infty$. In Figure~\ref{fig:distfuncs_Sec4}, we plot several examples of anyon gas distribution functions alongside the Fermi-Dirac and Bose-Einstein distributions. Notably, despite their superficial similarity to the Fermi-Dirac distribution, the anyonic distribution curves all differ in shape even after rescaling by $k$.

We emphasize that anyons always become strongly interacting as temperature is lowered, and the effects of their braiding statistics take on essential importance. Consequently, our anyonic distribution functions are strictly speaking only sensible in the dilute regime at high temperatures and low \emph{electron} densities, ${\rho_{\mathrm{electron}}=\rho_{\mathrm{anyon}}/k}$. In particular, for the abelian example, Eq.~\eqref{eq:abelian_finite_distr_fn_sec4}, we have neglected the  splitting of multi-particle states due to braiding phases discovered in Ref.~\cite{Arovas1985}, which are proportional to $1/k$. Hence, any extrapolation to lower temperatures or smaller values of $k$ is an  approximation. Outside of this regime, anyons are strongly interacting and their thermodynamics cannot be captured through the occupation statistics of free particles.

Before proceeding to compute observables and study further examples, we comment on the relationship between Eq.~\eqref{eq:abelian_finite_distr_fn_sec4} and other anyon gas distribution functions introduced in the past. Early proposals were based on Haldane's fractional exclusion statistics~\cite{Ramanathan1992,Wu1994-dg,Nayak1994}, where the Hilbert space dimension was reduced by a fractional value each time a state was filled. The resulting distribution functions possessed branch cuts and corresponded to partition functions with negative probabilities, due to approximations where state occupancies were treated independently. Although these negative probabilities were later understood and repaired~\cite{Murthy1999-rf}, this progress came at the cost of a properly defined one-particle distribution function. Our framework based on fusion constraints provides an alternate route: Although the fusion singlet coefficients defining physical states generally result in correlations between occupancies of different states, they factorize into products of quantum dimensions in the thermodynamic limit, Eq.~\eqref{eq:VerlindeAsymptotic}, meaning that the occupancies of different states effectively become independent, as in Eq.~\eqref{eq:single_level_part_fn}. While mixing of occupancies for different states is significant for few-anyon systems or outside of the dilute limit, the many-anyon systems we consider in this work admit an effective one-particle distribution function without need to invoke negative probabilities or fractional Hilbert space dimensions. See Appendix~\ref{app:Haldane_stats_appF} for a discussion of how to obtain our results from a generalization of the fractional statistics formalism in which the Hilbert space always has integer dimension.

On the other hand, Eq.~\eqref{eq:abelian_finite_distr_fn_sec4} evokes non-abelian counterparts obtained in the 't Hooft limit~\cite{Geracie2015-ya,Minwalla:2020ysu,Minwalla2022-pg}. This is not a coincidence, as we show in Section~\ref{sec:Examples_sec5} (with details in Appendix~\ref{app:tHooft}) that those results can also be reproduced in the occupation sequence formalism.


\subsection{High temperature expansion: The fermionic Laughlin gas}

In possession of the distribution functions, Eq.~\eqref{eq:distr_general_sec4} -- \eqref{eq:abelian_finite_distr_fn_sec4}, we can immediately compute thermodynamic observables. We focus on the example of a gas of $e/k$ anyons arising from doping a fermionic Laughlin state. This anyon gas is described by the modified Fermi-Dirac distribution, Eq.~\eqref{eq:abelian_finite_distr_fn_sec4}. Many of the lessons drawn from that case may also be applied to non-abelian examples, with physical quantities additionally depending on quantum dimensions. 

All thermodynamic quantities can be derived from the grand potential,
\begin{equation}\label{eq:free_energy}
    \Omega = -\frac{1}{\beta} \log Z_{\mathrm{ideal}} = -\frac{1}{\beta}\sum_i \log z_i \equiv - \frac{V}{\beta} \int_0^\infty d\epsilon ~ g(\epsilon) \log z(\epsilon)
\end{equation}
where $g(\epsilon)$ denotes the density of states per unit area. To compute concrete expressions, we must assume an anyon dispersion. We choose a non-relativistic dispersion, $\epsilon_{\bs{q}} = |\bs{q}|^2/2m_\star$, for a continuous wave vector, $\bs{q}=(q_x,q_y)$, leading to a constant density of states, ${g(\epsilon)\equiv g_0 = 2\pi m_\star}$.

The equation state for the abelian anyon gas takes the simple form,
\begin{equation}
    \varepsilon = p\,,
\end{equation}
irrespective of the length of the occupation sequence, $k$. Here $\varepsilon$ is the energy density and $p=-\Omega/V$ is the pressure. Formally exact expressions for the pressure, density, heat capacity per unit volume, and entropy density are presented in Appendix~\ref{app:abelian_thermo_appC}, alongside computations of their leading behavior in the low-temperature (Sommerfeld) and small $\mu/T$ expansions. Strictly speaking, however, the extrapolation of the partition function to temperatures below (or chemical potential above) $T\sim\mu(T=0)$ is uncontrolled, as the factorization ${Z\approx Z_{\mathrm{ideal}}\,Z_{\mathrm{int}}}$ generally breaks down. Nonetheless, some physical intuition may still be gleaned from those results -- such as the possibility of a Luttinger's theorem for anyons -- and we refer interested readers to Appendix~\ref{app:abelian_thermo_appC}.

Here we carry out the high-temperature expansion of the pressure, $p$, in powers of $\rho/g_0 T$, where $\rho$ is the \emph{anyon} number  density (see also Appendix~\ref{app:abelian_thermo_appC} for details of the calculation). Known as the virial expansion, it manifestly applies to the dilute regime for all finite values of $k$, and it tells us about the leading corrections to the classical ideal gas law. Famously, the virial expansion was carried out in Ref.~\cite{Arovas1985} for generic anyon systems without constraints on the anyon occupancy of physical states. There, the second virial coefficient was seen to possess sensitivity to the anyons' fractional statistics. In our dilute anyon models, where fusion constraints are paramount, we will see some notable deviations from the results of that work due to our emphasis on fusion.

We first expand the ratio, $p/\rho T$, in small fugacity, $\zeta = e^{\beta\mu}$. The result can be transparently organized in terms of corrections to the small-fugacity expansion for the Bose gas,
\begin{equation}
\label{eq: fugacity_k}
    \frac{p}{\rho T} = \left( \frac{p}{\rho T} \right)_\mathrm{Bose} + ~ \frac{k}{k+1} \zeta^k(\rho,T) + \mathcal{O}(\zeta^{k+1})\,,
\end{equation}
where the first term above represents the terms in the corresponding Bose gas expansion. 

To obtain an expansion in terms of the density, i.e. in powers of in powers of $\rho/g_0 T$, we solve self-consistently for the fugacity using the expansion,
\begin{equation}
    \rho = g_0 T \left( \zeta + \frac{\zeta^2}{2} + \frac{\zeta^3}{3} + \ldots + \frac{\zeta^k}{k} +\mathcal{O}(\zeta^{k+1})\right)\,.
\end{equation}
The first correction to the Bose gas virial expansion can then be computed by plugging the leading-order solution, ${\zeta=\rho/g_0T+\dots}$ into Eq.~\eqref{eq: fugacity_k}. This tells us that the first non-trivial correction to $p$ appears at order $\rho^{k+1}$, 
\begin{equation}\label{eq:virial_k}
    \frac{p}{\rho T}-\left( \frac{p}{\rho T} \right)_\mathrm{Bose} = \frac{k}{k+1} \left( \frac{\rho}{g_0 T} \right)^k+\mathcal{O}\left(\left[\frac{\rho}{g_0T}\right]^{k+1}\right)
\end{equation}
corresponding to the $(k+1)^{\mathrm{th}}$ virial coefficient, $B_{k+1}$,
\begin{align}\label{eq:B_k+1}
B_{k+1}-(B_{k+1})_{\mathrm{Bose}}&=\frac{k}{k+1}\,.
\end{align}
where $(B_{k+1})_{\mathrm{Bose}}$ is the corresponding Bose gas virial coefficient. For $k=1$, this result reproduces the correct second virial coefficient for the Fermi gas. Additionally, we note that all higher virial coefficients also receive $k$-dependent corrections. We therefore conclude that fusion constraints produce new corrections to the virial expansion, although they are insufficient to generate the correction to the second virial coefficient found in Ref.~\cite{Arovas1985}, which arises from braiding statistics and is sub-leading in $1/k$.

Eqs.~\eqref{eq:virial_k} -- \eqref{eq:B_k+1} admit a natural interpretation in terms of fusion constraints. The virial coefficient, $B_m$, is determined from the physics of $m$-body states. For a quantum gas of fermions, they encode Pauli exclusion statistics: The correction to the second virial coefficient is non-vanishing because no two fermions can occupy the same state. For this dilute anyon gas, on the other hand, up to $k$ anyons can fill a state, at which point they can fuse to an electron, which in turn \emph{must} obey Pauli exclusion. If one attempts to introduce another, $(k+1)^{\mathrm{th}}$, anyon, it will be pushed into a an open state. This is why it is the $(k+1)^\mathrm{th}$ virial coefficient that sees the first correction relative to the Bose gas. Another reason for the difference from Ref.~\cite{Arovas1985}, then, comes from our restricted Hilbert space of physical states. In that work, constraints descending from the presence of underlying microscopic electrons were not considered, and the braiding interactions of anyon pairs played the central role.

The above expansion was carried out in terms of the anyon density. Interestingly, deviations from the ordinary Bose gas virial expansion arise \emph{to all orders} if we expand in the \textit{electronic} pressure, $p_e = p/k$, and density, $\rho_e = \rho / k$. Now we find
\begin{equation}
    \frac{p_e}{\rho_e T} = 1 - \frac{k}{4} \frac{\rho_e}{g_0\, T} + \mathcal{O}\left( \frac{\rho^2}{g_0^2T^2} \right)\,.
\end{equation}
In particular, the corrections to the first $k$ virial coefficients are simply powers of $k$. If $\mathcal{B}_m$ are the virial coefficients of the ordinary Bose gas,
\begin{equation}
    \frac{p}{\rho_e T} = 1 + \sum_{m=1}^\infty B^e_{m+1} \left( \frac{\rho_e}{T} \right)^m, \qquad B_m^e = k^m \mathcal{B}_m, \quad m \le k\,.
\end{equation}
We view this expansion as the physically correct virial expansion, with ${\rho_e/T=\rho/kT}$ small (the physical electrons are dilute). This expansion allows one to take the $k\rightarrow\infty$ limit, where strength of braiding effects is weak, without the electron density, $\rho/k$,  vanishing. It would be of great interest to obtain corrections to the second virial coefficient~\cite{Arovas1985} by introducing braiding perturbatively in the context of this expansion. We leave this to future work.

Finally, the expansion in terms of electronic quantities allows us to naturally compute the high temperature entropy per electron in the virial expansion,
\begin{equation}
    \frac{S-S_{\mathrm{Bose}}}{N_e} = \log k + \mathcal{O}\left( T^{-1} \right)\,,
\end{equation}
where $S_{\mathrm{Bose}}$ is the high-temperature entropy of a Bose gas with density $\rho$, and ${N_e=\rho\,(\mathrm{Area})/k}$. Hence, the constant correction to the total entropy at high temperatures counts the number of anyons fusing to an electron.


\subsection{Multi-anyon mixtures}

Thus far, we have focused on gases populated by a single anyon species. But almost any realistic topological order will possess multiple anyon types, each affiliated with its own energy spectrum. The single-anyon assumption is then only justified at temperatures below the gap of the next-lightest anyon, Eq.~\eqref{eq:temp_for_single_flavour_gas_sec3}. As the temperature is increased while holding the chemical potential fixed, this next-lightest anyon, which we will call $b_1$, will be thermally activated at $T\sim \Delta_\text{next-lightest}$, with the average number of $b_1$ anyons in the gas vanishing exponentially at lower temperatures. As before, we continue to assume that the gap of the topological order is the largest scale, in particular $\Delta_\text{next-lightest} < \Delta_\mathrm{TO}$.

Our occupation sequence formalism can be straightforwardly adapted to such multi-anyon mixtures. If $\mu < \Delta_\text{next-lightest}$, the anyon gas should remain dilute, and it will remain primarily comprised of the lightest anyon, $a_1$, such that inter-species interactions are negligible. Concretely, we consider the regime,
\begin{equation}
    \Delta_0 < \mu < \Delta_\text{next-lightest}, \qquad \Delta_\text{next-lightest} \lesssim T \ll \Delta_\text{next-next-lightest}\,,
\end{equation}
where we have a two-species mixture of anyons $a_1$ and $b_1$, while other anyons are thermally suppressed. Each can be associated with its own occupation sequence,
\begin{equation}
    \begin{split}
        \ell_a &= (1,a_1, a_2, \ldots)\,,\\
        \ell_b &= (1,b_1, b_2, \ldots)\,.
    \end{split}
\end{equation}
Under the dilute approximation, each anyon species is treated as non-interacting. The ideal anyon gas partition function therefore separates into factors associated with each\footnote{The situation is somewhat more subtle if $b_1$ is a fusion outcome of $a_1$ occurring along its occupation sequence, such that it may be realized as a $n$-fold bound state of $a_1$'s. Then a sufficient condition for Eq.~\eqref{eq:multi_anyon_partition_factorization_sec4.5} to hold is that $n\,\Delta_{\mathrm{lightest}}<\Delta_{\mathrm{next-lightest}}$. We expect this condition to hold generically, since the local particles appearing on $\ell_a$ should be lighter than those along $\ell_b$.},
\begin{equation}\label{eq:multi_anyon_partition_factorization_sec4.5}
    Z_{\mathrm{ideal}} = Z[\ell_a]\, Z[\ell_b]\,,
\end{equation}
where we introduce the notation, $Z[\ell]$, as the ideal partition function constructed from the sequence $\ell$.

All thermodynamic observables can thus be obtained additively, with the entropy and heat capacity in particular given by
\begin{equation}
    \begin{split}
        s &= s[\ell_a] + s[\ell_b]\,,\\
        c_V &= c_V[\ell_a] + c_V[\ell_b]\,.
    \end{split}
\end{equation}
One physical consequence of these relations is that, as the temperature is increased above $\Delta_{\mathrm{next-lightest}}$, the heat capacity should jump by an amount dictated by the next-lightest anyon. We therefore expect the heat capacity in anyon gases to exhibit \emph{thermally activated steps} analogous to those in ordinary, un-fractionalized systems with gapped degrees of freedom. For example, consider a gas of $e/3$ quasiparticles doped into a $\nu=1/3$ Laughlin state: As the temperature is raised above the gap to the charge-$2e/3$ quasiparticle, the heat capacity in Eq.~\eqref{eq:high_T_abelian_thermo_sec4} (for $k=3$) will experience a jump.


\section{Examples: Fermi seas and Bose condensates from anyon gases}
\label{sec:Examples_sec5}

Having introduced our general treatment of anyon gas thermodynamics, here we consider several more concrete examples based on doping different abelian and non-abelian topological orders, with local particles that are either bosons or fermions. In each case, we start by postulating a lightest anyon and constructing an occupation sequence, which in turn leads to a distribution function from which observable quantities can be computed. As we saw in the examples of abelian anyon gases studied in Section~\ref{sec:distr_fn_occupation_sequence_sec4}, anyon gases assembling into local fermions will produce Fermi-Dirac-like metallic distribution functions, while those assembling into local bosons can produce analogues of the Bose-Einstein distribution, signalling the onset of anyon superfluidity or superconductivity at low temperatures. Intriguingly, in the latter cases, only non-abelian examples show deviations from the Bose-Einstein distribution.     


\subsection{Anyon superfluids}

We start with gases of anyons that fuse into local bosons, which can arise in two distinct situations. First, one may consider doping charge into a topological order emerging from a system of microscopic, charge-$e$ bosons, such as a $\nu=1/2$ bosonic Laughlin state. The constituents of the resulting anyon gas can only fuse into bosonic local particles, and Bose condensation at low temperatures will be the typical outcome. This is the traditional scenario for anyon superfluidity~\cite{Laughlin1988,Fetter1989,Lee1989,Chen1989,Fradkin1990a}. On the other hand, starting from a topological order of microscopic, charge-$e$ fermions, one can imagine doping anyons that either must fuse to bosonic bound states or (in non-abelian examples) prefer to do so energetically. For example, charge-$2e/3$ quasiparticles doped into a $\nu=1/3$ fermionic Laughlin state can only yield charge-$2e$ local bosons, i.e. Cooper pairs~\cite{Shi2024doping}. Alternatively, in a non-abelian anyon gas, fusion channels yielding local bosons may be energetically preferred even if the parent topological order is fermionic~\cite{Shi2025-nj}. Below, we will see each of these scenarios can be accounted for in our formalism through the appropriate choice of occupation sequence.


\subsubsection{Bosonic Laughlin states}\label{sec:Bose_Laughlin_4.1.1}

Bosonic Laughlin states are associated with even denominator filling, $\nu =1/2m$, with $m>0$ a positive integer. The anyon content of these theories can be expressed in terms of a UMTC whose anyons form cyclic groups, $\mathbb{Z}_{2m}$,
\begin{equation}
    \mathcal{C}_{\nu=1/2m} = \{ 1, a, a^2, \ldots, a^{2m-1}\}, \qquad a^{2m} = 1\,.
\end{equation}
We remind the reader that $1$ denotes a local boson. Suppose we pick $a_1 = a$ to be the lightest particle. Then two occupation sequences are possible, as discussed in Section~\ref{sec:distr_fn_occupation_sequence_sec4}. The first repeats indefinitely, with local bosons all allowed to occupy the same quantum state,
\begin{equation}\label{eq:SeqBoseLaughlin}
    \ell_{\nu=1/2m} = (1,a,a^2,\ldots,a^{2m-1},1,a,a^2\ldots)\,.
\end{equation}
The other possible sequence truncates at the first local boson, which one may associate with an energetic constraint. The physics of this case is analogous to the fermionic sequences discussed in Section~\ref{sec:distr_fn_occupation_sequence_sec4} and correspond to the distribution function in Eq.~\eqref{eq:abelian_finite_distr_fn_sec4}. For our purposes here, we focus on the infinitely repeating sequence. 

Because the bosonic Laughlin state is abelian, the quantum dimensions of all anyons are unity. The single-level partition function thus reads
\begin{align}
    z_{\nu=1/2m}(\epsilon) &= \sum_{n=0}^{\infty} e^{-n\beta(\epsilon-\mu)}=\frac{1}{1-e^{-\beta(\epsilon-\mu)}}.
\end{align}
This is exactly the partition function of a Bose gas, and thermodynamic observables are insensitive to the presence of fractionalization, in striking contrast to situations with fermionic local particles, as in Eq.~\eqref{eq:abelian_finite_distr_fn_sec4}. We will soon see that deviations from the Bose-Einstein distribution require the presence of non-abelian anyons.


\subsubsection{A comment on duality and the semion gas}

Topological orders can have many dual representations, and anyon gases are no different. Among the standard dualities of topological orders are so-called \emph{level-rank dualities}, which formally relate topological quantum field theories (TQFTs) described by ordinary gauge fields to those described by gauge fields that are spin$_c$ connections, which can couple to fermionic degrees of freedom (see Ref.~\cite{Senthil:2018cru} for an accessible introduction). Roughly, these dualities equate different forms of flux attachment: One should equally well be able to describe a topological order with composite fermions coupled to a spin$_c$ gauge field as with composite bosons coupled to an ordinary gauge field. How do these dualities manifest in the statistical mechanics of anyon gases?

Perhaps the simplest level-rank duality applies to the semion gas. As was understood many years ago~\cite{Zhang1989}, one may describe a semion with $\pi/2$ statistics by attaching ``one half'' of a flux quantum to a boson. In field theoretic language, this is achieved by coupling a bosonic Landau-Ginzburg theory to a $\mathrm{U}(1)_{-2}$ Chern-Simons gauge field. An equivalent description of semions starts with fermions but instead attaches ``minus one half'' flux quanta, which shifts their statistics as $\pi-\pi/2=\pi/2$. This procedure corresponds to a field theory of fermions coupled to a $\mathrm{U}(1)_{+2}$ spin$_c$ gauge field (and different couplings to the background electromagnetic field). Taken together, these facts indicate that the $\mathrm{U}(1)_{-2}$ TQFT is equivalent to the $\mathrm{U}(1)_2$ spin$_c$ TQFT.

Since we have already presented the outcome of our analysis for the ordinary formulation of the semion gas, it remains to consider its spin$_c$ dual. Formally, this involves introducing an additional ``transparent'' fermion, $\psi$, that has trivial mutual statistics with all other anyons. The resulting spin$_c$ semion theory then consists of the following anyons,
\begin{equation}
\label{eq: spinc semion umtc}
    \mathcal{C}_\mathrm{semion} = \{ 1, a, \psi, a_\psi \}, \qquad a^2 = 1\,.
\end{equation}
where $a_\psi$, the semion, is the composite anyon obtained from merging together $a$, which we term the anti-semion, and the local fermion $\psi$. It may be loosely understood as the outcome of flux attachment to the fermion. 

As elaborated on in Appendix~\ref{app:UMTC}, Eq.~\eqref{eq: spinc semion umtc} is not a UMTC. However, the physical topological order is obtained by gauging fermion parity: The transparent fermion is not the physical electron, since the microscopic particles are all bosons. Indeed, all the correct fusion coefficients are obtained from the quotient of this theory by fermion parity, leading to a UMTC consisting only of two anyons,
\begin{equation}
    \mathcal{C}_\mathrm{semion} / \mathbb{Z}_2 = \{ 1, b \}, \qquad b^2 = 1\,.
\end{equation}
where $b$ is now a collective placeholder for either the semion $a$ or its fermion-twisted cousin $a_\psi$. This quotiented presentation is equivalent to the bosonic UMTC studied in the previous subsection.

To see the equivalence explicitly, suppose now that we pick $a_1 = a_\psi$ as the lightest anyon and construct the following occupation sequence,
\begin{equation}
    \ell_\mathrm{semion} = ( 1, a_\psi, 1, a_\psi, \ldots )\,.
\end{equation}
Its quotiented cousin is 
\begin{equation}
    \ell_\mathrm{semion} / \mathbb{Z}_2 = ( 1, b, 1, b, \ldots )\,,
\end{equation}
which is equivalent to the occupation sequence in the previous subsection. We therefore trivially find the following expression for the partition function,
\begin{equation}
    z_\mathrm{semion}(\epsilon) = \frac{1}{1-e^{-\beta(\epsilon-\mu)}}\,.
\end{equation}
Once again we find a Bose condensate. Importantly, because the transparent fermion is not present in the quotiented UMTC, we could not have arrived at it with a physical occupation sequence. Although this example may appear trivial, the analysis here can be adapted to much more elaborate non-abelian level-rank dualities, which generally relate $\mathrm{SU}(N)_k$ spin$_c$ TQFTs to ordinary $\mathrm{U}(k)_{-N}$ TQFTs, as we explain in Appendix~\ref{app:tHooft}.


\subsubsection{The $2e/3$ gas}\label{sec:2/3_Laughlin_superfluid_5.1.3}

Bose condensates can also arise on doping a topological order emerging from microscopic fermions, such as a ${\nu=2/3}$ -- or its particle-hole conjugate, ${\nu=1/3}$ -- Laughlin state, provided the lightest anyon fuses to an even-charge bound state, such as a charge-$2e$ Cooper pair. Indeed, for a charge-$2e/3$ anyon gas, the local particles are Cooper pairs arising by fusing three anyons. This intuition has been behind some recent proposals for anyon superconductivity proximate to the $\nu=2/3$ FQAH state~\cite{Shi2024doping,Shi2025plateau,nosov2025}, and it is also naturally accounted for via a judicious choice of occupation sequence. 

We start with a somewhat expanded view of the $\nu=2/3$ Laughlin state that accommodates the possibility of electronic bound states and is nicely compatible with the (usually bosonic) UMTC formalism. The anyon content of this state is given by the following list,
\begin{equation}\label{eq:Laughlin_2/3_anyon_set_sec5.1.3}
    \mathcal{C}_{\nu=2/3} = \{ 1, a, a^2, e , a^4, a^5 \}, \qquad a^3 = e, \quad (a^2)^3 = e^2= 1\,.
\end{equation}
where $e$ denotes the microscopic electron, which is a local fermion. The anyon $a$ is the fundamental anyon of charge $e/3$, meaning each $a^p$ has charge $p\,e/3$. The Cooper pair of charge-$2e$ is denoted by the identity, $1$. Note the anyons $a^4$ and $a^5$ can be equally obtained by respectively fusing $a$ and $a^2$ with the electron, $e$.

If $a^2$ is the lightest anyon, then the gas obtained on doping the $\nu=2/3$ state is comprised primarily of charge-$2e/3$ anyons, which cannot form a local electron by fusing among one another. The only alternative is to form the Cooper pair, $1=(a^2)^3$, a local boson. We see this happen naturally from the occupation sequence point of view. Excluding the possibility that local bosons are energetically disallowed from occuping the same state, the remaining occupation sequence is again infinite,
\begin{equation}
    \ell_{ 2e/3} = ( 1, a, a^2, 1, a, a^2, 1, \ldots )\,,
\end{equation}
for which the partition function reduces to that of a regular superfluid
\begin{equation}
    z_{2e/3}(\epsilon) = \frac{1}{1-e^{-\beta(\epsilon-\mu)}}\,,
\end{equation}
and the distribution function is the Bose-Einstein distribution. Hence, the tendency toward condensation of Cooper pairs is immediate within in our formalism. However, we emphasize that the Bose-Einstein distribution above should not be taken too literally. As $T_c$ is approached, interactions will become important and round the Bose-Einstein condensate into a superconductor. Indeed, from the point of view of the dilute anyon gas at high temperature, we cannot even assess to whether the resulting anyon superconductor is in the weak (BCS) or strong (BEC) pairing regime. In upcoming work, we will present a theory of the anyon gas applicable to the normal state close to $T_c$.


\subsubsection{The Bose-Fibonacci gas}
\label{Sec4-bosonic-fib}

To study properties of non-abelian anyon gases, we choose the simplest example, which also happens to be most salient for its potential in topological quantum computation. Consider a Fibonacci topological order, with a single non-abelian anyon, $\tau$, that can fuse either to itself or to a microscopic boson, again denoted $1$. Its UMTC data consists of 
\begin{equation}
    \begin{split}
        \mathbf{Fib} &= \{ 1, \tau \}\,,\\
        \tau \times \tau &= 1 + \tau\,,\\
        d_\tau &=  \frac{1+\sqrt{5}}{2}\,.
    \end{split}
\end{equation}
Because the local particles are bosons, Bose condensation is again a natural outcome, and our interest is in infinitely repeating occupation sequences. 

However, as discussed in Section~\ref{sec:distr_fn_occupation_sequence_sec4}, non-abelian anyon gases are richer than abelian ones: their occupation sequences cannot be chosen uniquely and depend on energetic differences between fusion channels. In fact, in this example, the number of distinct occupation sequences is \emph{infinite}! For example, one could choose $(1,\tau,1,\dots)$ or $(1,\tau,\tau,1,\dots)$ or $(1,\tau,\tau,\tau,1,\dots)$, and so on. Furthermore, unlike abelian topological orders, each of these occupation sequences will lead to a distinct variant of the Bose-Einstein distribution, albeit each with universal behavior in the $T\rightarrow\infty$ limit (at least in this example). 

For concreteness, we consider the simplest sequence,
\begin{equation}\label{eq:Fib}
    \ell_\mathrm{Fib} = ( 1, \tau, 1, \tau, \ldots )\,.
\end{equation}
We expect that this sequence may generically have some energetic preference, since the bosons formed possess minimal charge, whereas other sequences produce higher-charge boson bound states. 

\begin{figure}
\centering
\begin{minipage}[b]{0.5\columnwidth}
    \centering
    \includegraphics[width=1\columnwidth]{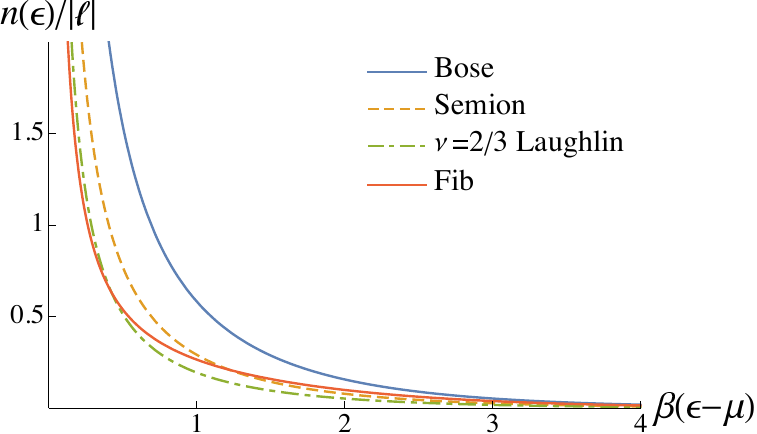}   
\end{minipage}
\caption{Bose condensate distribution functions, rescaled to represent the occupation of microscopic bosons. Doping abelian topological orders produces Bose-Einstein distributions, up to rescaling by the occupation sequence length. On the other hand, doping non-abelian topological orders yields additive corrections, altering the shape of the distribution function at low temperatures.}\label{fig:DistBose}
\end{figure}

Using the formula~\eqref{eq:single_level_part_fn}, we compute the associated single-state partition function, 
\begin{equation}
    z_\mathrm{Fib}(\epsilon) = \sum_{n=0}^\infty\Big( e^{-2 \beta n (\epsilon - \mu)} + d_\tau\, e^{-\beta (2n+1) (\epsilon - \mu)}\Big)= \frac{1 + d_\tau\, e^{-\beta(\epsilon-\mu)}}{1-e^{-2\beta(\epsilon-\mu)}}\,,
\end{equation}
which in turn yields the distribution function,
\begin{equation}
\label{eq:Bose-Fib-dist}
    n_\mathrm{Fib}(\epsilon) = \frac{1}{d_\tau^{-1}\, e^{\beta(\epsilon - \mu)} + 1} + \frac{1}{e^{2\beta(\epsilon-\mu)}-1}\,.
\end{equation}
We compare this distribution function to its abelian counterparts in Figure~\ref{fig:DistBose}. Notably, like our results for anyon gases that fuse to local fermions, this distribution function shares properties of both the Bose-Einstein and Fermi-Dirac distributions, with the additive Fermi-Dirac-like contribution arising from the presence of a non-trivial quantum dimension. In this case, however, the ultimate low-temperature physics favors Bose condensation below a critical temperature.

To our knowledge, while modified Fermi-Dirac distributions have been proposed for anyon gases in the past~\cite{Ramanathan1992,Wu1994-dg,Nayak1994,Geracie2015-ya,Minwalla:2020ysu,Minwalla2022-pg}, this is the first time a modified Bose-Einstein distribution has been found. This is a testament both to the generality of the occupation sequence concept, as well as to the non-trivial fusion properties of non-abelian anyon gases. It would be interesting to explore the landscape of different non-abelian distribution functions for theories with more elaborate fusion structures, although we leave this to future work. Finally, we note other recent work that has also considered thermodynamic properties of Fibonacci topological orders from other vantages~\cite{Hu2013-nl,Ritz-Zwilling2024}.

\subsection{A non-abelian anyon metal}

We now return to examples exhibiting Fermi-Dirac statistics, signalling a tendency toward metallic behavior at low temperatures (although, in the presence of attractive interactions, a superconducting instability is another possibility). In particular, if an anyon gas is produced on doping a fermionic topological order, such ``anyon metals'' are always formally possible~\footnote{Strinctly speaking, the true $T\rightarrow 0$ phase will either be a composite Fermi liquid metal or a Fermi liquid of local electrons. We use the term ``anyon metal'' to reference the finite-temperature dynamics giving way to these phases. This is analogous to how the terms ``anyon superfluid'' and ``anyon superconductor'' tend to be used to refer to the dynamics of the normal state, even though the ultimate ground state phase may be completely conventional.}. They correspond to an occupation sequence where the lightest anyon ultimately fuses to a local fermion. As mentioned above, anyon gases based on bosonic topological orders can have similar physics if the occupation sequence is truncated by hand, enforcing a Pauli-like occupation constraint energetically. However, the low temperature physics of such states should be more akin to correlated insulators or Wigner crystals -- with charges becoming localized due to interactions -- and should not be regarded as metallic. While those examples may be of some physical interest, here we choose to focus on the fermionic examples.

For fermionic Laughlin states, we obtained the corresponding distribution function in  Eq.~\eqref{eq:abelian_finite_distr_fn_sec4} and already considered its observable properties in detail. What remains is to consider how the physics changes if one instead dopes a non-abelian topological order. Here we again choose to focus on the paradigmatic example of a Fibonacci anyon gas, albeit this time arising from a system of local fermions. For a review of the details of this topological order, including a trial wave function, see Ref.~\cite{Goldman2020}.

The fermionic extension of the Fibonacci topological order has anyon content,
\begin{equation}
    \mathbf{Fib}_\psi = \{ 1, \tau, \psi, \tau_\psi \}\,,
\end{equation}
where $\psi$ is the local fermion (the microscopic electron), and we use the notation ${\tau_\psi=\tau\times\psi}$ to denote a second species of Fibonacci anyon obtained on fusing with the local fermion. The non-trivial fusion rules are
\begin{equation}
    \tau \times \tau = 1 + \tau, \qquad \tau \times \tau_\psi = \psi + \tau_\psi, \qquad \tau_\psi \times \tau_\psi = 1 + \tau\,.
\end{equation}
The quantum dimensions of the two Fibonacci anyons are the same,
\begin{equation}
    d_\tau = d_{\tau_\psi} = \frac{1+\sqrt{5}}{2}\,.
\end{equation}

\begin{figure}[t]
\centering
\begin{minipage}[b]{0.5\columnwidth}
    \centering
    \includegraphics[width=1\columnwidth]{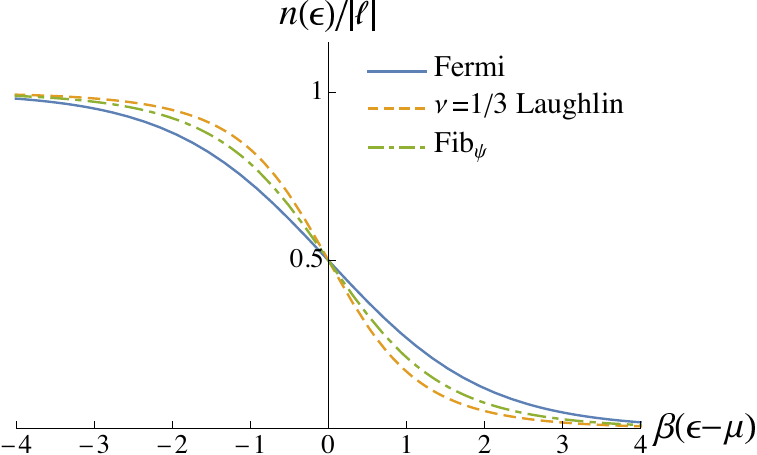}   
\end{minipage}
\caption{Distribution functions for anyonic Fermi seas, featuring the $e/3$ Laughlin gas, the Fermi-Fibonacci gas, and the ordinary Fermi-Dirac distribution. All curves are re-scaled by the occupation sequence length to indicate the number of local particles. 
}\label{fig:DistFermi}
\end{figure}

We choose the lightest anyon to be $\tau_\psi$. As in the case of the Bose-Fibonacci gas, there are an infinite number of possible occupation sequences, each corresponding to a different number of Fibonacci fusion outcomes prior to reaching the local fermion, $\psi$, that truncates the sequence. For simplicity, we again choose the sequence of minimum length,
\begin{equation}
    \ell_{\mathrm{Fib}_\psi} = ( 1, \tau_\psi, \tau, \psi )\,,
\end{equation}
Plugging this sequence into Eq.~\eqref{eq:single_level_part_fn}, we find a single-level partition function and a distribution function,
\begin{align}
\label{eq:Fermi-Fib-partition}
    z_{\mathrm{Fib}_\psi}(\epsilon) &= 1 + d_{\tau} e^{-\beta(\epsilon-\mu)} + d_{\tau} e^{-2\beta(\epsilon-\mu)} + e^{-3\beta(\epsilon-\mu)}\,, \\
\label{eq:Fermi-Fib-distribution}
    n_{\text{Fermi-Fib}}(\epsilon) &= \frac{1}{z_{\mathrm{Fib}_\psi}}
            \left( d_{\tau} e^{-\beta(\epsilon-\mu)} + 2d_{\tau} e^{-2\beta(\epsilon-\mu)} + 3e^{-3\beta(\epsilon-\mu)} \right).
        \end{align}
Like its abelian counterpart in Eq.~\eqref{eq:abelian_finite_distr_fn_sec4}, this result resembles the Fermi-Dirac distribution, but again has a different shape (see Figure~\ref{fig:DistFermi}). From Eqs.~\eqref{eq:Fermi-Fib-partition} -- \eqref{eq:Fermi-Fib-distribution}, it is natural to interpret the non-trivial quantum dimensions as shifting the effective temperatures felt by different anyon types.

We emphasize that, in contrast to their abelian counterparts, the choice of lightest anyon for a non-abelian state does not uniquely determine whether the system tends to form a Fermi sea or a Bose condensate. The ultimate outcome is determined by the energetic hierarchy of different fusion channels. For example, the infinite  sequence $(1,\tau_\psi,1,\tau_\psi,\dots)$ is also allowed and would correspond to an anyon superfluid at low temperatures. We already computed its Bose-Einstein-like distribution function given in Eq.~\eqref{eq:Bose-Fib-dist}. A closely related example, in the context of the Moore-Read state, is presented in Ref.~\cite{Shi2025-nj}.


\subsection{Exactly solvable large-$N$ models}

The discussion thus far has relied on approximate descriptions of anyon gases valid in dilute limits. Luckily, in some notable cases it is possible to make progress starting from exactly solvable models. For example, recent studies of string-nets~\cite{Levin2004-gq} deformed from solvability~\cite{Vidal:2021isf,Ritz-Zwilling2024} have presented results for thermodynamic observables for anyon gases at zero chemical potential by following a complementary formalism to ours.

As discussed in Section~\ref{Sec2}, Chern-Simons-matter theories can model the fluctuations of itinerant anyons by coupling a TQFT to a matter sector comprised of fermions or bosons, with gauge fluctuations transmuting the matter particles into anyons. Although in general these theories are very strongly coupled at low energies, an exactly solvable large-$N$ limit exists and is known as the 't Hooft limit. Like Fibonacci gases, all anyons in the 't Hooft limit are non-abelian, yet nevertheless a wide range of physical properties of these anyon gases can be computed \emph{exactly}, leading to a vast literature computing everything from the thermal free energies~\cite{Giombi2011-xo,Ofer2012-thermal}, to transport~\cite{Geracie2015-ya,Gur-Ari2016-zc}, to scattering amplitudes~\cite{Mehta2022-xt}, to quantum oscillations in a background magnetic field~\cite{Halder:2019foo}. One reason for these successes is that in the 't Hooft limit anyon gases remain dilute \emph{for all values of the density}, i.e. the scale at which the dilute approximation breaks down is suppressed by $1/N$~\cite{Minwalla:2020ysu,Minwalla2022-pg} (see also Appendix~\ref{app:diluteness}).

In Appendix~\ref{app:tHooft}, we show explicitly that distribution functions obtained in the 't Hooft limit can be naturally reproduced using our anyon statistical mechanics formalism. We outline the basic argument here. 

The anyon content of $\mathrm{SU}(N)_k$ Chern-Simons theory is furnished by irreducible representations of the group $\mathrm{SU}(N)$. Starting from the fundamental representation, $F$, which is an $N$-component vector, and its complex conjugate $\overline{F}$, also known as the anti-fundamental representation, further representations can be generated by taking through taking products. For instance, we can generate the symmetric and anti-symmetric representations from
\begin{equation}
    F \times F = S_2 + A_2\,,
\end{equation}
where $S_n$ and $A_n$ respectively denote the symmetric and antisymmetric representations with $n$ tensor indices. The singlet ($1$) and adjoint ($\mathrm{Adj}$) representations can be generated from
\begin{equation}
    F \times \overline{F} = 1 + \mathrm{Adj}\,.
\end{equation}
Further products with $F$ or $\overline{F}$ give rise to more irreducible representations. The number of symmetric and anti-symmetric indices in any representation are respectively capped at $k$ and $N$, resulting in a finite number of irreducible representations and thus a finite number of anyons.

The 't Hooft large $N$ limit is a double scaling limit:
\begin{equation}
    k, N \rightarrow \infty, \qquad \lambda = \frac{N}{k + \sgn(k) N} = \mathrm{fixed}
\end{equation}
where $\lambda$, called the \textit{'t Hooft coupling}, ranges from $0$ (decoupled) to $1$ (strongly coupled). Note that the denominator of $\lambda$ depends on a choice of UV regularization for the Chern-Simons theory, as the $\mathrm{SU}(N)_k$ Chern-Simons level is well known to receive a quantum shift the presence of a nonzero Yang-Mills term~\cite{ChenSemenoffWu1992}. For the purposes of the analysis in this section, we choose to place this shift into the definition of $\lambda$ in order to match with the chosen regularization in Refs.~\cite{Minwalla:2020ysu,Minwalla2022-pg}.

The choice of lightest anyon in our framework corresponds to the choice of matter type (boson or fermion) and $\mathrm{SU}(N)$ representation in Chern-Simons-matter theory. For example, if the lightest anyon is in the representation $F$, then the multi-anyon states are constrained by whether the matter is bosonic or fermionic, resulting in symmetric or anti-symmetric representations respectively. We can now immediately construct the occupation sequences,
\begin{equation}\label{eq:CSM_sequences_sec5}
    \begin{split}
        \ell_\mathrm{CSB} &= ( 1, F, S_2, S_3, \ldots, S_k )\,,\\
        \ell_\mathrm{CSF} &= ( 1, F, A_2, A_3, \ldots, A_N )\,.
    \end{split}
\end{equation}
Here the subscripts CSB and CSF refer to whether the matter fields are bosons or fermions. We show in Appendix~\ref{app:tHooft} that these sequences give the exact same partition function as what was computed in Refs.~\cite{Minwalla:2020ysu,Minwalla2022-pg} using 't Hooft's large $N$ expansion.  It is very interesting that only the sequences in Eq.~\eqref{eq:CSM_sequences_sec5} are picked out in the 't Hooft limit, which suppresses all other fusion patterns energetically (such as those leading to infinite sequences and thus Bose condensation).

An astute reader will have noticed that the bosonic sequence in Eq.\ \eqref{eq:CSM_sequences_sec5} does not satisfy the conditions we impose on the occupation, owing to the fact that the maximal symmetric representation $S_k$ is not a local boson or fermion. The maximal antisymmetric representation $A_N$ on the other hand, is indeed a local fermion (boson) for odd (even) $N$. We believe the resolution to be that in the strict $N,k\rightarrow\infty$ limit, this isn't an issue because the sequence is simply infinite, but we emphasize that this is no longer the case for any finite $N,k$, highlighting the potential role of subleading corrections to the large-$N$ limit.


\section{Discussion}
\label{Sec6}

Starting from universal fusion data along with a minimal set of dynamical assumptions, we have shown that it is possible to construct a statistical mechanics for anyon gases based on doping \emph{any} topological order. Our models apply to temperature regimes where the anyons can be considered dilute, and long-ranged interactions can be neglected, allowing for an effective single-particle description in the thermodynamic limit. 

The possibility of such a dilute regime may come as a surprise to many due to anyons' highly entangled nature; it certainly breaks down at low temperatures and for few-anyon systems, where occupation of different states become correlated. However, our formalism incorporates the key data for fractionalization through an anyonic exclusion principle, which amounts to a requirement that physical anyonic states must be related to electronic states through fusion. This requirement, along with assumptions about the anyon spectrum encoded in an occupation sequence, allows us to write down large families of distribution functions for anyon gases that interpolate between the Bose-Einstein and Fermi-Dirac distributions, with the tendency toward Bose condensation or the formation of a Fermi sea being ultimately determined by the lightest local particle in the spectrum.

We anticipate that our framework will provide a launching point for more detailed studies of anyon dynamics at finite temperature. For example, armed with anyonic distribution functions, it should be possible to study finite-temperature transport properties of anyon gases through the development of classical and quantum Boltzmann equations. Such work promises to make crucial predictions relevant to recent experiments in 2$d$ moir\'{e} materials, where it is necessary to elucidate anyon dynamics' role in producing a panoply of exotic correlated phases. It also may enable us to push beyond the dilute regime, including long-ranged interactions and braiding at the level of a generalized collision integral. We plan to pursue this direction in upcoming work.

We also remark that lattice symmetry of real FQAH systems should act projectively on the anyons~\cite{Lu-Ran-Oshikawa_2020}. While we have neglected symmetry enrichment in this work, we believe our formalism should be directly applicable to such settings through assigning global symmetry quantum numbers to the anyons. We also leave this direction for the future.

Finally, our work complements and extends earlier studies of anyon thermodynamics~\cite{Arovas1985} from the point of view of fractional exclusion principles~\cite{Haldane1991-ln,Murthy:1994yew,Ramanathan1992,Wu1994-dg,Nayak1994,Schoutens:1997xh,Ardonne2001-wx}, large-$N$ field theories~\cite{Geracie2015-ya,Minwalla:2020ysu,Minwalla2022-pg}, and solvable lattice models~\cite{Vidal:2021isf,Ritz-Zwilling2024}. In particular, it would be interesting to study whether aspects of fractional exclusion statistics that are hidden in our approach emerge through finite corrections to the asymptotic formula in Eq.~\eqref{eq:VerlindeAsymptotic}. We also expect that a fruitful direction would be to adapt our framework based on TQFT ideas to quantum Hall matrix models~\cite{Susskind2001,Polychronakos2001}, which have a direct connection to Calogero-Sutherland models~\cite{Morariu:2005pv}, where fractional exclusion statistics can be studied exactly.


\section*{Acknowledgements}

 We are especially grateful to Seth Musser for collaboration at early stages of this work. We also thank Fiona Burnell, Dominic Else, Eduardo Fradkin, Zohar Komargodski, Ho Tat Lam, Sri Raghu, Shinsei Ryu, Zhengyan Darius Shi, Raman Sohal, D. T. Son, and Carolyn Zhang for helpful discussions and comments on the manuscript. YN is supported by the Funai Overseas Scholarship. HG is supported by startup funds at the University of Minnesota. UM is supported by a Simons Postdoctoral Fellowship in Ultra-Quantum Matter as well as a Gordon and Betty Moore Foundation Grant GBMF10279.


\begin{appendix}


\section{Dilute limits for $\mathrm{SU}(k)_N$ Chern-Simons-matter theories}
\label{app:diluteness}


In this Appendix, we present a more detailed and general argument for the existence of a dilute window for anyon gases, based on a dual formulation that has the mean-field behavior of a (non-)Fermi liquid. This description enables one to more directly consider the evolution to the dilute regime at temperatures of order the density. This evolution is complicated if approached from the Landau level picture presented in Section~\ref{Sec2}, as it involves both Landau level mixing from short-ranged interactions and thermal activation of matter particles between Landau levels.

\subsection{Model}

Start with a gas of charge-$e/k$ anyons doped into the $\nu=1/k$ Laughlin state, which can be described as bosons, $\phi$, coupled to a $\mathrm{U}(1)_{-k}$ Chern-Simons gauge field~\cite{Zhang1989}. Like its fermionic cousin described in Section~\ref{Sec2}, at zero temperature these bosons fill $k$ Landau levels. At $T=0$, and for $k$ even, they form a bosonic IQH state~\cite{Levin2013}, which on integrating out the bosons yields an anyon superfluid. For $k$ odd, they instead form a metallic state analogous to the well known bosonic composite Fermi liquid at $\nu=1$. These states are referred to as ``secondary composite Fermi liquids'' in Ref.~\cite{Shi2025-nj}. Despite the matter degrees of freedom being bosons -- whose short-ranged interactions are crucial for stabilizing the aforementioned zero temperature phases -- on heating this system well above the bosons' cyclotron gap, ${T\gg\omega_c=\rho_\phi/(k\,m_\star)}$, these interactions can be ignored. One again expects Debye screening of gauge fluctuations, and the argument in Section~\ref{Sec2} again applies. 

Now we consider an equivalent formulation. The anyon gas described by $\mathrm{U}(1)_{-k}$ Chern-Simons theory coupled to bosons can also be represented in terms of non-relativistic fermions at finite chemical potential, coupled to a $\mathrm{SU}(k)_1$ gauge field. This duality\footnote{Strictly speaking, this duality is only meant to hold for the ground state at $T=0$. Nevertheless, here we assume it extends at least to the temperature scale of the leading irrelevant operator, which here is the Yang-Mills term.} follows the level-rank duality that equates the $\mathrm{U}(1)_{-k}$ theory to $\mathrm{SU}(k)_1$ with spin$_c$ connection~\cite{Aharony2016-uf,Hsin2016blu}. Here, despite the gauge group being non-abelian, the anyons are all abelian: Fluctuations of cherge-$e/k$ Laughlin quasiparticles are understood in terms of the fermionic matter fields in the fundamental representation, while the remaining anyons are naturally obtained through fusion and thus can be viewed as multi-fermion bound states. 

Without affecting the essential physics, we can further generalize this picture to the non-abelian case of a $\mathrm{SU}(k)_N$ gauge theory coupled to fermions, which is dual to $\mathrm{U}(N)_{-k}$ coupled to bosons. Concretely, we introduce non-relativistic, fermionic ``quark'' fields, $\chi_\alpha$, $\alpha=1,\dots,k$, coupled to a $\mathrm{SU}(k)_{N}$ gauge field at finite chemical potential, $\mu$. In Euclidean spacetime, $(\tau=it,\boldsymbol{x})$,
\begin{align}
\label{eq:non-abelian gas}
S&=S_\chi+S_{\mathrm{CS}}+S_{\mathrm{YM}} \\
\label{eq:Schi}
S_\chi&=\int d\tau\,d^2\boldsymbol{x}\left(\chi^\dagger D_\tau \chi-\frac{1}{2m_\star}|D_i \chi|^2-\mu\, \chi^\dagger_\alpha \chi_\alpha\right)\,,\qquad D_\mu =\partial_\mu-ib_\mu\,,\\
S_{\mathrm{CS}}&=\int d\tau\,d^2\boldsymbol{x}\,\frac{iN}{4\pi}\Tr\left[bdb-\frac{2i}{3}b^3\right]\,,\\
S_{\mathrm{YM}}&=\int d\tau\,d^2\boldsymbol{x}\,\frac{k}{2g_{\mathrm{YM}}^2}\Tr[f^2]\,,
\end{align}
where $b_\mu=b_\mu^a t^a_{\alpha\beta}$, $t^a$, are the $k^2-1$  generators of $\mathrm{SU}(k)$, and $D_\mu\chi=\partial_\mu\chi-ib^a t^a\chi$. We have assumed a parabolic dispersion, and $m_\star$ is the fermion effective mass. We also have defined the field strength, $f = db - i b\wedge b=D_{[\mu}b_{\nu]}$ ($D_\mu$ here being the covariant derivative acting on the adjoint representation). We use $i,j,\dots=x,y,\dots$ to denote  spatial indices. Note we assume vanishing external magnetic field, as would be the case were one obtain this theory by doping a FQAH state. We also neglect any discrete lattice symmetry and assume a circular Fermi surface.

The model in Eq.~\eqref{eq:non-abelian gas} describes a Fermi sea of the quarks, whose density is ${\rho_\chi=\langle\chi^\dagger\chi\rangle}$. At low energies, the quarks are strongly coupled to the gauge field, $b_\mu$, forming a non-Fermi liquid metal. Our interest will be in temperature scales well above the onset of non-Fermi liquid behavior but below the scale of the Yang-Mills term, $g_{YM}^2/k$, corresponding to the scale where gauge fluctuations propagate as decoupled gluons. In this special regime, the gauge field dynamics are frozen out, as its physics is controlled by the Chern-Simons part of the action. Its only role is to constrain the physical Hilbert space of the matter variables to fusion singlets, and the partition function may be approximated as that of a non-interacting gas of anyons.

\subsection{Scaling argument for diluteness}

To see this argument play out explicitly, we start by integrating out the fermions to obtain a functional determinant,
\begin{align}
Z&=\int\mathcal{D}b\,\det\left[D_\tau+\frac{1}{2m_\star}D_i^2-\mu\right]\exp\left(-S_{\mathrm{gauge}}\right)\,\\
&=\int\mathcal{D} b\,\exp\left(-S_{\mathrm{CS}}-S_{\mathrm{YM}}-W\right)\,,
\end{align}
where $W=\Tr\log[D_\tau+D_i^2/2m_\star-\mu]$. Expanding $W$ in powers of $b$ gives a Hertz-Millis effective potential, which in momentum and imaginary frequency space can be expressed to quadratic order as
\begin{align}
W[b]&=\int \frac{d\Omega\, d^2\boldsymbol{q}}{(2\pi)^3}\Tr\left[b_\mu(-i\Omega,-\bs{q})\,\Pi_{\mu\nu}(i\Omega,\bs{q})\, b_\nu(i\Omega,\bs{q})+\mathcal{O}(b^4)\right]\,.
\end{align}
This expansion is well known to be an uncontrolled description of the $T\rightarrow 0$ physics and possesses IR singularities~\cite{Halperin1993,Abanov2004,Lee:2009epi}. Nevertheless, it suffices for our humbler goal of identifying the temperature scale at which quantum corrections are sub-leading compared to the terms in $S_{\mathrm{CS}}$, such that the gauge field dynamics are frozen out. 

For simplicity, we adopt the non-abelian analogue of Coulomb gauge, $\partial_i b_i=0$. The leading (one-loop) effect of fermion fluctuations is overdamping of the transverse components of the gauge field ($a_T\equiv \varepsilon_{ij}q^ia^j(\bs{q})/|\bs{q}|$) and screening of the scalar fluctuations ($a_\tau$). Suppressing color indices, the diagonal components of the polarization tensor, $\Pi_{\mu\nu}$, are
\begin{align}
\Pi_{\tau\tau}(i\Omega,\boldsymbol{q})=\frac{K_F}{2\pi v_F}\left(1-\frac{\Omega}{\sqrt{\Omega^2+|\boldsymbol{q}|^2}}\right)\,&,\qquad\Pi_{TT}(i\Omega,\bs{q})=\frac{v_F K_F}{2\pi}\frac{\Omega}{\sqrt{\Omega^2+|\boldsymbol{q}|^2}}\,.
\end{align}
where $\Pi_{TT}$ denotes the transverse spatial component of the polarization,  ${K_F=\sqrt{2m_\star\mu}\propto\sqrt{\rho_\chi}}$ is the Fermi wave vector, $v_F=K_F/m_\star$ is the bare Fermi velocity, and we have suppressed color indices. To reduce notational clutter, we will set  $v_F=1$ in the analysis below.

We now turn on a finite temperature $T$ and compare terms in the quadratic gauge boson action. For this purpose, we can imagine replacing $\Omega$ with $T$ -- this will lead to the same conclusions as writing out the sums over bosonic Matsubara frequencies, $\Omega_n=2\pi n\, T$, in full. We start by comparing the one-loop corrections in $\Pi_{\mu\nu}$ to the Yang-Mills term, $k\Tr[f^2]/g_{YM}^2$. Ignoring the Chern-Simons term, at temperatures above
\begin{align}
T_{LD}=\sqrt{\frac{K_F g_{YM}^2}{2\pi\,k}}\,,
\end{align}
the Yang-Mills term dominates over quantum corrections. In this regime, the singular gauge fluctuations are relativistic, with $\Omega_n\sim T\sim |\bs{q}|$. On the other hand, below $T_{LD}$, the singular gauge fluctuations have $\Omega_n\ll |\bs{q}|$ and are purely transverse, with the Landau damping term $\Pi_{TT}\sim (K_F/2\pi)\,\Omega_n/|\bs{q}|$ dominating the frequency-dependent term $\sim k\,\Omega^2_n/g_{YM}^2$ in the Yang-Mills contribution of the gauge dispersion. 

We now introduce the Chern-Simons term, the frequency-dependent part of which goes like ${N\Omega_n\sim NT}$. For this term to dominate \emph{both} the contribution from the polarization, $\sim K_F/(2\pi k)$, \emph{and} the Yang-Mills term, $\sim kT^2/g_{YM}^2$, the temperature should satisfy
\begin{align}
\frac{K_F}{2\pi k}\sim\frac{\sqrt{\rho_\chi}}{k}\ll NT\,&,\qquad \frac{k}{g_{YM}^2}T^2\ll NT\,,
\end{align}
where we have again used the fact that at sufficiently high temperatures the gauge fluctuations are relativistic, $\Omega_n\sim |\bs{q}|$. Dividing out by temperature, we obtain the final inequality relating charge density, temperature, and the Yang-Mills coupling,
\begin{align}
\label{eq: SU(N)k diluteness}
\frac{\sqrt{\rho_\chi}}{kN}\ll T\ll \frac{N}{k}g_{YM}^2\,.
\end{align}
We have therefore arrived at the diluteness condition for the anyon gas described by $\mathrm{SU}(k)_N$ gauge theory coupled to fermions at finite density. In dilute temperature window, Eq.~\eqref{eq: SU(N)k diluteness}, the gauge fluctuations have essentially no dynamics of their own and are governed by the Chern-Simons TQFT action. Above this window, the gauge field propagates as in $\mathrm{SU}(N)$ Yang-Mills theory with linear dispersion, while below it the transverse gauge fluctuations are overdamped and eventually lead to non-Fermi liquid effects in all their usual glory/infamy. 

For $N=1$, Eq.~\eqref{eq: SU(N)k diluteness} resembles the condition obtained in the main text  for the abelian theory. Indeed, reintroducing the Fermi velocity, $v_F=K_F/m_\star$, the left-hand-side of Eq.~\eqref{eq: SU(N)k diluteness} can be replaced with $(1/kN)(\rho_\chi/m_\star)$, although we emphasize that the effective mass $m_\star$ is non-universal and completely unrelated to that in the abelian example considered in the main text. If the limit $k\rightarrow\infty$ is taken with $kT$ held fixed, at low densities there should be a finite (but small) temperature window where the diluteness condition is satisfied. 

We also remark that the diluteness condition becomes exact \emph{for all densities} in the 't~Hooft limit, where $k\rightarrow\infty,N\rightarrow\infty$ and $k/N$ is held fixed. This is a consequence of the $1/k$ suppression of $\Pi_{\mu\nu}$, which pushes the scale of interaction effects like the overdamping of the gauge field, as well as of quasiparticle braiding, down to $T=0$. Although we have carried out this analysis for the particular case of $\mathrm{SU}(N)_k$ Chern-Simons theory coupled to fermions, recent exact  calculations~\cite{Minwalla:2020ysu,Minwalla2022-pg} indicate that this aspect of our conclusion should hold for any 't Hooft limit Chern-Simons-matter theory.

\subsection{Consequences of diluteness and partition function factorization}

The scaling argument above establishes the existence of a dilute window for anyon gases obtained by doping $\mathrm{SU}(k)_N$ Chern-Simons-fermion theories. Here we argue that in this temperature window the thermal partition function can be approximated by the ``ideal'' partition function in Eq.~\eqref{eq:partition-fnc-factorization}.

We return to the thermal $\mathrm{SU}(k)_N$ Chern-Simons-fermion action,
\begin{equation}\label{eq:CSM_action_expanded_appA}
    \begin{split}
        S &= \int_0^\beta d\tau\, \int d^2\boldsymbol{x}\, \chi^\dagger \left( D_\tau + \frac{1}{2m_\star} D_i^2 - \mu \right) \chi\\
        &- i \int_0^\beta d\tau\, \int d^2\boldsymbol{x}\, b^a_\tau \left( \tilde{\rho}_\chi^a - \frac{N}{2\pi} f^a_{xy} \right)\\
        &+ \frac{iN}{4\pi}\int_0^\beta d\tau\, \int d^2\boldsymbol{x}\, \Tr \left[ b_x \partial_\tau b_y \right]\,.
    \end{split}
\end{equation}
where $f_{12}^a$ is the $SU(k)$ magnetic field strength, and $\tilde{\rho}_\chi^a=\chi^\dagger t^a\chi$ is the non-abelian ``color'' density, which we distinguish from $\rho_\chi=\chi^\dagger\chi$, the physical EM charge density.

Whereas earlier we first integrated out the $\chi$-fermions, now we integrate first over $b_\tau$. In the dilute regime, the Yang-Mills term is unimportant compared to the Chern-Simons term, meaning that $b_\tau$ can be taken as a Lagrange multiplier enforcing Gauss' law,
\begin{equation}
\label{eq:constraint_app}
    \tilde{\rho}_\chi^a = \chi^\dagger t^a\chi = \frac{N}{2\pi} f_{xy}^a\,,
\end{equation}
as a quantum operator constraint. The resulting partition function is
\begin{align}
    Z \approx \int \mathcal{D}\chi\, \mathcal{D}b_i\, \delta\left( \rho_\chi^a - \frac{N}{2\pi} f_{xy}^a \right) \exp\left(-S_\chi^0+a_i\, j_\chi^i+\frac{iN}{4\pi}\int_0^\beta d\tau\int d^2\bs{x}\Tr \left[ b_x \partial_\tau b_y \right]\right)\,.
\end{align}
where $S_\chi^0$ is the free fermion action and ${j_\chi^i=\frac{i}{2m_\star}(\chi^\dagger t^a\partial^i\chi-\partial^i\chi^\dagger t^a\chi)}$ is the fermion gauge current. Now integrating over $\chi$ again yields a constrained functional determinant, \emph{taken only over states satisfying Eq.}~\eqref{eq:constraint_app}.

What are the consequences of this? The remaining Chern-Simons term in the action for $b_i$ tells us that the conjugate momentum to $b_x$ is $b_y$ and \emph{vice-versa}, leading to the standard Chern-Simons commutation relations and enforcing their Wilson lines' status as $\operatorname{SU}(k)_N$ anyons with the necessary fusion and braiding properties. This may be thought of as a non-abelian generalization of flux attachment. The delta function in the path integral therefore requires $\chi$-quanta to obey the fusion rules of Wilson lines in pure Chern-Simons theory, as well as disallowing multi-particle states that do not fuse to Chern-Simons singlets under these modified fusion rules in order to enforce gauge invariance.

Furthermore, our previous scaling analysis allows us that in the dilute regime the path integral over $b_i$ can be treated using a saddle-point approximation, in a manner analogous to the analysis in Ref.~\cite{Ofer2012-thermal}, and its fluctuations can be neglected to leading order in $1/k$. 
Including fluctuations in $b_i$, which mediate anyon-anyon interactions, will lead to corrections to the energies of multi-anyon states, which in principle one can compute to reproduce the corrections to the anyon energy spectrum seen in Ref.~\cite{Arovas1985}.

The thermal partition function can then be reduced simply to a sum over flux sectors of $b_i$ (i.e. anyon labels), weighted by fusion singlet coefficients,
\begin{equation}
    Z = \sum_{\{n_k\}} \mathcal{N} \left( \{A_{n_k}\} \right) e^{-\beta\sum_k n_k(\epsilon_k - \mu)}\,.
\end{equation}
where $n_k \ge 0$ are occupation numbers for momenta $k$, $\epsilon_k$ is the free fermion dispersion, $A_{n_k}$ is the anti-symmetric combination of $n_k$ fundamental representations of $\mathrm{SU}(k)$, and $\mathcal{N}$ is the fusion-singlet coefficient in pure Chern-Simons theory.


\section{Computing fusion coefficients}
\label{app:UMTC}

\subsection{Commutative fusion categories}

We derive in this appendix the asymptotic formula for the fusion-singlet coefficient in bosonic and fermionic topological orders. For both cases we don't require the full mathematical structure of a (unitary or super) modular tensor category, but instead just a commutative fusion category, i.e., a set of anyons with fusion rules
\begin{equation}\label{eq:fusion_rules_appB}
    a\times b = \sum_c N_{ab}^c c, \qquad N_{ab}^c = N_{ba}^c
\end{equation}
In any commutative fusion category, the fusion matrix $\boldsymbol{N}_a$ with components
\begin{equation}
    (\boldsymbol{N}_a)_{bc} \equiv N^c_{ba}
\end{equation}
can be diagonalized by a unitary \textit{mock $S$-matrix} $\tilde{S}_{ab}$ \cite{bonderson2007non,Barkeshli2019} with the eigenvalues $\lambda^{(a)}_i$ of $\boldsymbol{N}_a$ given by
\begin{equation}
    \lambda^{(a)}_i = \frac{\tilde{S}_{ai}}{\tilde{S_{1i}}}
\end{equation}
This can be used to derive the Verlinde formula
\begin{equation}
    N_{ab}^c = \sum_d \frac{\tilde{S}_{ad} \tilde{S}_{bd} \tilde{S}^*_{cd}}{\tilde{S}_{1d}}
\end{equation}
where $1$ is the trivial boson. For bosonic topological order, the mock $S$-matrix is the same as the modular $S$-matrix which measures the linking number of two anyon loops. Fermionic theories, however, do not have a modular $S$-matrix but the mock $S$-matrix still exists. The matrix form of the fusion coefficients are useful for computing higher order fusion coefficients, since
\begin{equation}
    a_1 \times \ldots \times a_n = \sum_b ( \boldsymbol{N}_{a_2} \boldsymbol{N}_{a_3} \cdots \boldsymbol{N}_{a_n} )_{a_1 b} ~ b
\end{equation}
where the term in the parenthesis is the matrix multiplication of the fusion matrices. Using the diagonalization of the fusion matrices we can then write the following
\begin{equation}\label{eq:exact_fusion-singlet_coeff_appB}
    \mathcal{N}_g(a_1,\ldots,a_n) \equiv (\boldsymbol{N}_{a_2} \cdots \boldsymbol{N}_{a_n})_{a_1 1} = \sum_b \tilde{S}_{b1}^{2-2g} \prod_{i=1}^n \frac{\tilde{S}_{b a_i}}{\tilde{S}_{b1}}
\end{equation}
Now, it's possible to show that $\tilde{S}_{a1} = d_a / \mathcal{D}$ \cite{bruillard2017}, where $d_a$ is the quantum dimension of $a$ and $\mathcal{D}$ is the total quantum dimension. For UMTCs this is straightforward to see, since $\tilde{S}=S$ is the linking number of two anyon loops normalized by the total quantum dimension, while $d_a$ is the expectation value of a single loop.

If the number $n$ of fusing anyons is much larger that the number of anyon types, Equation \eqref{eq:exact_fusion-singlet_coeff_appB} can be simplified using a saddle point approximation which picks out only the largest eigenvalues of the fusion matrices $\boldsymbol{N}_{a_i}$, which are the quantum dimensions $d_{a_i}$. With this result, we have the following asymptotic formula for the fusion-singlet coefficient
\begin{equation}
    \mathcal{N}_g(a_1,\ldots,a_n) \simeq \mathcal{D}^{2-2g} \prod_{i=1}^n d_{a_i}
\end{equation}


\subsection{Including fermions in the fusion-singlet coefficient}

Our derivation of the fusion-singlet formula so far has only been for the case when the fusion result is the local boson. We can also show that the same asymptotic form applied for the fusion to a local fermion, whenever possible.

Recall from Section~\ref{Sec3} that in fermionic theories there exists a second transparent line $\psi$ that braids trivially with every anyon and obeys the following fusion rule
\begin{equation}
    \psi \times \psi = 1
\end{equation}
It is distinct from the local boson $1$ in that it has non-trivial topological spin $\theta_\psi = -1$ and as such, generates a fermion parity symmetry $\mathbb{Z}_2^f$. An anyon theory that includes a local fermion is called a \textit{super-modular tensor category} (sMTC). Importantly, sMTCs are not UMTCs in that they do not furnish a unitary representation of the modular group, and thus the mock $S$-matrix that diagonalizes the fusion matrix $\boldsymbol{N}_a$ is not the same as the modular $S$-matrix. However, the results of the previous section still hold for the mock $S$-matrix.

The presence of fermion parity allows us to construct a smaller fusion category of the sMTC $\mathcal{C}$ by quotienting the trivial fermionic theory $\mathcal{C}_\psi = \{ 1, \psi \}$, which we will label $\mathcal{C}/\psi$. This \textit{fermionic quotient} theory consists of anyons modulo fusion with the local fermion, i.e., topological supercharges of the form $\hat{a} = (a,a_\psi)$ where $a\in\mathcal{C}$. In particular, the quotient has one unique transparent line denoted by $\hat{1} = (1,\psi)$ instead of two. The fusion coefficient of the quotient take the same values as those of the sMTC up to consistency with fermion parity
\begin{equation}
    N_{\hat{a}\hat{b}}^{\hat{c}} = N_{ab}^c = N_{a_\psi b_\psi}^c = N_{a_\psi b}^{c_\psi} = N_{ab_\psi}^{c_\psi}
\end{equation}
The same is true for the quantum dimensions
\begin{equation}
    d_{\hat{a}} = d_a = d_{a_\psi}
\end{equation}
The fusion-singlet coefficient of the quotient can then be described in terms of fusion coefficients of the sMTC in the following way
\begin{equation}
    \begin{split}
        \mathcal{N}_g(\hat{a}_1, \ldots, \hat{a}_n) &= N_{a_1 \ldots a_n}^1 + N_{a_1\ldots a_n}^\psi\\
        N_{a_1\ldots a_n}^1 &= (\boldsymbol{N}_{a_2}\cdots\boldsymbol{N}_{a_n})_{a_1 1}\\
        N_{a_1\ldots a_n}^\psi &= (\boldsymbol{N}_{a_2}\cdots\boldsymbol{N}_{a_n})_{a_1 \psi}
    \end{split}
\end{equation}
where the $a_i$ can be chosen to be any of the two anyons corresponding to the topological supercharge $\hat{a}_i$. The fusion-singlet coefficient of the quotient defines the fusion-singlet that we require to be non-zero for a physical state in a fermionic theory. This coefficient has an expression in terms of the mock $S$-matrix (dropping the tilde for notational simplicity) $\hat{S}_{\hat{a}\hat{b}}$ of the quotient:
\begin{equation}
    \mathcal{N}_g(\hat{a}_1,\ldots,\hat{a}_n) = \sum_{\hat{b}} \hat{S}_{\hat{b}\hat{1}}^{2-2g} \prod_{i=1}^n \frac{\hat{S}_{\hat{b}\hat{a}_i}}{\hat{S}_{\hat{b}\hat{1}}}
\end{equation}
The mock $S$-matrix of the quotient may in general be distinct from that of the sMTC, but in the large $n$ limit the asymptotic form still holds owing to the equality of the fusion matrices and quantum dimensions of the sMTC and its quotient, and we have
\begin{equation}
    \mathcal{N}_g(\hat{a_1},\ldots,\hat{a}_n) \simeq \mathcal{D}^{2-2g} \prod_{i=1}^n d_{a_i}
\end{equation}
and we are free to use the same asymptotic formula irrespective of whether we have fermionic or bosonic topological order.


\section{Thermodynamics of the fermionic Laughlin gas}\label{app:abelian_thermo_appC}

\subsection{Exact equation of state}
We present exact expressions for the partition function and various thermodynamic quantities for the femrionic Laughlin gas described by the occupation sequence
\begin{equation}
    \ell_k = (1, a_e, a^2, \ldots, e)\,,
\end{equation}
where $k$ is an odd number, $e$ is the electron, and $a_e$ is the charge $e/k$ quasiparticle, with subscript denoting fusion with the electron. The partition function per level is a finite geometric series, and the corresponding distribution is given by
\begin{equation}
    \begin{split}
        z_k(\epsilon) &= \frac{y^{m+1}-1}{y-1}\,,\\
        n_k(\epsilon) &= \frac{y}{y-1} \left( k - (k+1) \frac{y^k-1}{y^{k+1}-1} \right)\,,
    \end{split}
\end{equation}
where $y = e^{-\beta(\epsilon-\mu)}$. Assuming a constant density of states $g(\epsilon) = g_0$, the pressure takes the following form
\begin{equation}
    \begin{split}
        p &= - \frac{1}{\beta} \int_0^\infty d\epsilon ~ \log z(\epsilon)\\
        &= g_0 T^2 \left[ \mathrm{Li}_2\left(\zeta\right) - \frac{1}{k+1} \mathrm{Li}_2\left(\zeta^{k+1}\right) \right]\,,
    \end{split}
\end{equation}
where $\zeta = e^{\beta\mu}$ is the fugacity and $\mathrm{Li}_2(z) = - \int_0^z du \frac{\log(1-u)}{u}$ is the dilogarithm function, which has a branch cut on the positive real axis for $|z|>1$. The difference of the two dilogarithms kills the imaginary part as long as the branch choice is the same for both, making the pressure well-defined everywhere on the complex plane, including the branch cut. Note that if we drop the second dilogarithm in the pressure, we get the expression for the pressure of the Bose gas.

Thermodynamic quantities can be computed by taking suitable derivatives of the grand potential, Eq.~\eqref{eq:free_energy}. The energy density, $\varepsilon$, is exactly the same as the pressure, making the equation of state of the gas independent of $k$,
\begin{equation}
    \varepsilon = p\,.
\end{equation}
The entropy density, heat capacity per volume, and particle-number density are given by
\begin{equation}
    \begin{split}
        \rho &= g_0 T ~ \log \frac{\zeta^{k+1}-1}{\zeta-1} \,, \\
        c_V = s &= g_0 T^2 \left[ 2 T \left( \mathrm{Li}_2\left(\zeta\right) - \frac{1}{k+1} \mathrm{Li}_2 \left(\zeta^{k+1}\right) \right) - \mu \log\frac{\zeta^{k+1}-1}{\zeta-1} \right]\,.
    \end{split}
\end{equation}


\subsection{Virial expansion}

The virial expansion is obtained by first expanding thermodynamic potentials in powers of fugacity and the solving for the fugacity order by order in powers of the density. This is achieved straightforwardly by using the series form of the dilogarithm function:
\begin{equation}
    \mathrm{Li}_2(z) = \sum_{j=1}^\infty \frac{z^j}{j^2}
\end{equation}
as well as the series expansion for $\log(1-z)$ after writing the density as
\begin{equation}
    \rho = - g_0 T \left[ \log\left( 1-\zeta \right) - \log\left( 1 - \zeta^{k+1} \right) \right]\,.    
\end{equation}
We find the following expressions:
\begin{equation}
    \begin{split}
        p &= g_0 T^2 \left( \zeta + \frac{\zeta^2}{4} + \frac{\zeta^3}{9} + \ldots + \frac{\zeta^{k+1}}{(k+1)^2} + \ldots - \frac{\zeta^{k+1}}{k+1} - \frac{\zeta^{k+2}}{4(k+1)} - \ldots \right)\,,\\
        \rho &= g_0 \left( \zeta + \frac{\zeta^2}{2} + \frac{\zeta^3}{3} + \ldots + \frac{\zeta^{k+1}}{k+1} + \ldots - \zeta^{k+1} - \frac{\zeta^{k+2}}{2} - \ldots \right)\,.
    \end{split}
\end{equation}
We've highlighted the term proportional to $\zeta^{k+1}$ above since that is the leading power of the fugacity whose coefficient differs from that of a Bose gas. Dividing the two and expanding the result in powers of the fugacity, we obtain the following:
\begin{equation}\label{eq:fugacity_exp_appC}
    \frac{p - p_\mathrm{Bose}}{\rho T} = \frac{k}{k+1} \zeta^k + \mathcal{O}\left(\zeta^{k+1}\right)\,.
\end{equation}
To turn this into a virial expansion we only need to solve for the fugacity order by order in the density. Up to order $\zeta^k$, this amounts to solving the following equation:
\begin{equation}
    \frac{\rho}{g_0\, T} = \zeta + \frac{\zeta^2}{2} + \ldots + \frac{\zeta^k}{k} + \mathcal{O}\left(\zeta^{k+1}\right)\,,
\end{equation}
which is identical to the corresponding equation of a Bose gas, since the leading correction coming from $\log(1-\zeta^{k+1})$ begins at the next order. Evaluating the fugacity order by order we immediately obtain the virial expansion
\begin{equation}
    \frac{p-p_\mathrm{Bose}}{\rho T} = \frac{k}{k+1} \left( \frac{\rho}{g_0\,T} \right)^k + \mathcal{O}\left( \frac{\rho^{k+1}}{T^{k+1}} \right)
\end{equation}

For the simplest case of the $\nu = 1/3$ Laughlin gas, we present the fugacity and virial expansions below:
\begin{equation}
    \begin{split}
        p &= g_0 T^2 \left[ \zeta + \frac{\zeta^2}{4} + \frac{\zeta^3}{9} - \frac{3\zeta^4}{16} + \mathcal{O}\left(\zeta^5\right) \right]\,,\\
        \rho &= g_0 T \left[ \zeta + \frac{\zeta^2}{2} + \frac{\zeta^3}{3} - \frac{3\zeta^4}{4} + \mathcal{O}\left(\zeta^5\right) \right]\,,\\
        p &= \rho T \left[ 1 - \frac{\zeta}{4} - \frac{7}{72} \zeta^2 + \frac{25}{36} \zeta^3 + \mathcal{O}\left(\zeta^4\right) \right]\,,\\
        &= \rho T \left[ 1 - \frac{1}{4} \left( \frac{\rho}{g_0 T} \right) + \frac{1}{36}\left( \frac{\rho}{g_0 T} \right)^2 + \frac{3}{4} \left( \frac{\rho}{g_0 T} \right)^3 + \mathcal{O}\left( \frac{\rho^4}{T^4} \right) \right]\,.
    \end{split}
\end{equation}


\subsection{Small chemical potential expansion}

We presented the virial expansion in powers of $\rho/T$ in Section~\ref{Sec3}. Here we present instead an expansion of the thermodynamic potentials in powers of the chemical potential. Expanding in powers of $\mu/T$, we obtain
\begin{equation}\label{eq:high_T_abelian_thermo_sec4}
    \begin{split}
        \varepsilon = p &= g_0\,T^2 \left[ \frac{k}{k+1}\frac{\pi^2}{6} + \log(k+1)\, \frac{\mu}{T} + \frac{k}{4}\frac{\mu^2}{T^2} + \mathcal{O}\left(\mu^3\over T^3\right) \right]\,,\\
        \rho &= g_0\,T \left[ \log(k+1)  + \frac{k}{2}\frac{\mu}{T} + \mathcal{O}\left(\mu^2\over T^{2}\right) \right]\,,\\
        c_V = s &= g_0\,T \left[ \frac{k}{k+1} \frac{\pi^2}{3}  +  \log(k+1)\,\frac{\mu}{T} + \mathcal{O}\left(\mu^2\over T^{2}\right) \right]\,,
    \end{split}
\end{equation}
where $\rho$ is the \emph{anyon} number density, $s$ is the entropy density, and $c_V$ is the heat capacity per unit volume. Intriguingly, despite the semiclassical nature of the high-temperature limit, \emph{all} terms in the series expansion depend on the fractional charge through $k$.

From the expression for the density in Eq.~\eqref{eq:high_T_abelian_thermo_sec4}, we immediately see that the dilute approximation holds self-consistently at fixed $\mu$,
\begin{equation}
    \frac{\rho}{g_0\,k} \ll   T\,.
\end{equation}
This is particularly apparent at large $k$, owing to the fact that $\log(k+1) \ll k$ for $k\gg 1$. The fact that the numerical values of the coefficients in the expansion are small suggest that the dilute approximation should hold even for finite values of $k$.

The entropy is also of special interest. In the thermodynamic limit, the entropy per local electron, $S/N_e$, is constant and determined by the occupation sequence length,
\begin{align}
\frac{S}{N_e}=\frac{s}{\rho/k}\sim2\,\log(k+1)\,,\qquad \textrm{as }N_e\rightarrow\infty\,.
\end{align}
That the entropy should depend on $k$ can be intuited from the parent $\nu=1/k$ Laughlin state giving rise to the anyon gas, whose  topological degeneracy on the torus is $k$.

This result for the entropy hints at a more general structure that applies even to non-abelian anyon gases. For a general dilute anyon gas with occupation sequence ${\ell=(1,a_1,\dots,a_{|\ell|})}$, the total entropy, $S$, which is a sum over entropies per state, $\varsigma(\epsilon)$, loosely admits an expression of the Shannon form,
\begin{equation}\label{eq:S-ln}
    \begin{split}
       S=\int d\epsilon\, g(\epsilon)\,\varsigma(\epsilon)\,,\qquad \varsigma(\epsilon) &= - \sum_{n=0}^{|\ell|} d_{a_n}\frac{e^{-n\beta(\epsilon - \mu)}}{z(\epsilon)} \log \left[ \frac{e^{-n\beta(\epsilon - \mu)}}{z(\epsilon)} \right]\\
        &\sim - \Tr \hat\rho \log \hat\rho\,,
        \end{split}
\end{equation}
where $z(\epsilon)$ is the single-state partition function of Eq.~\eqref{eq:single_level_part_fn}. The effective density matrix $\hat\rho$ -- not to be confused with the anyon density, $\rho$, which does not possess a hat -- would have eigenvalues $e^{-\beta n(\epsilon-\mu)}$ for all values of $n$ from $0$ to $|\ell|$, with (generally non-integer) ``degeneracies,'' $d_{a_n}$. Equation~\eqref{eq:S-ln} therefore suggests that at temperatures where the chemical potential becomes vanishingly small, the entropy per local electron, $S/N_e$, should go like the logarithm of the sum of quantum dimensions along the occupation sequence,
\begin{align}
\frac{S}{N_e}\sim\log d_{\mathrm{tot}}\,,\qquad d_{\mathrm{tot}}=\sum_{n=0}^{|\ell|}d_{a_n}\,.
\end{align}
The dependence of entropy on quantum dimensions is natural given their interpretation as anyon degeneracies~\cite{Yang2009-hf}.


\subsection{Low temperature expansion}

The high temperature expansions above   are in accordance with the dilute limit. We present here the low temperature expansions, valid for $T\ll\mu$, which all truncate for a constant density of states.
\begin{equation}
    \begin{split}
        \varepsilon = p &= g_0 \left( \frac{k}{2}\mu^2 + \frac{k}{k+1} \frac{\pi^2}{2} T^2 \right) + \mathcal{O}(e^{-\mu/T})\,,\\
        c_V = s &= g_0\, \frac{k}{k+1} \frac{2\pi^2}{3} T + \mathcal{O}(e^{-\mu/T})\,,\\
        \rho &= g_0\, k\, \mu\, + \mathcal{O}(e^{-\mu/T})\,.
    \end{split}
\end{equation}
All corrections to these are non-perturbatively small in $T/\mu$. The value of the pressure at vanishing temperature is the anyonic version of the degeneracy pressure. The entropy and heat capacity are linear in temperature as expected, and the entropy obeys the third law of thermodynamics. The value of the density is suggestive of an ``anyonic Luttinger's theorem'' relating the density to the chemical potential. Note in particular that this expansion is in contradiction with diluteness, which would require
\begin{equation}
    \frac{\rho}{g_0\, k} \ll T \qquad \implies \mu \ll T\,,
\end{equation}
so these expressions aren't necessarily applicable to a physical system. Rather they are simply illustrative of the consistency of the partition function with a Hilbert space interpretation.


\section{Non-abelian Chern-Simons-matter theories in the 't Hooft limit}
\label{app:tHooft}

Chern-Simons-matter theories are quantum field theories describing itinerant anyons. We will restrict ourselves to $\mathrm{SU}(N)$ gauge groups and consider two classes of theories distinguished by a fermionic or bosonic matter sector. The actions are given by
\begin{equation}\label{eq:CSM_actions_app}
    \begin{split}
        S_\mathrm{CSF}[\psi,a] &= \int dt d^2\boldsymbol{x} \sum_{j=1}^N \overline{\psi}_j ( i \slashed{D}_a - m ) \psi_j + S_\mathrm{CS}[a]\\
        S_\mathrm{CSB}[\phi,a] &= \int dt d^2\boldsymbol{x} \sum_{j=1}^N \left( |D_a \phi_j|^2 - m^2 \overline{\phi}_j \phi_j \right) + S_\mathrm{CS}[a] + \ldots\\
        S_\mathrm{CS}[a] &= \frac{k'}{4\pi} \int dtd^2\boldsymbol{x} ~ \Tr\left( ada + \frac{2}{3} a^3\right)
    \end{split}
\end{equation}
where $\psi$ and $\phi$ are vectors living in the fundamental (i.e., vector) representation of the gauge group\footnote{This action is rather subtle, since it obscures various counterterms and background gauge fields that enforce the existence of a local fermion. See, for example \cite{Hsin2016blu}}. We will denote by $k$ the level of the Chern-Simons theory that describes the ground state of these actions. The precise value of $k'$ above, in terms of $k$, depends on the choice of regularization \cite{ChenSemenoffWu1992}. We restrict our focus to theories that live in the duality web  and always have an underlying spin structure, so that there exists a local fermion in the ground state TQFT, which we will denote $c$ \cite{Hsin2016blu,Senthil:2018cru}.


\subsection{Anyon content}

The anyons of the spin-enhanced $\mathrm{SU}(N)_k$ Chern-Simons theory are related to irreducible representations of $\mathrm{SU}(N)$, and can be generated from the fundamental representation denoted by $F$, and its charge conjugate, the anti-fundamental, $\overline{F}$. The fusion rules for these two look like the following:
\begin{equation}
    \begin{split}
        F \times F &= S_2 + A_2\\
        F \times \overline{F} &= 1 + \mathrm{Adj}\\
        \overline{F} \times \overline{F} &= \overline{S}_2 + \overline{A}_2
    \end{split}
\end{equation}
where $S_2$ ($A_2$) is the two-index symmetric (anti-symmetric) representation, and $\mathrm{Adj}$ is the adjoint representation\footnote{We emphasize here that the symmetric, anti-symmetric, singlet, and adjoint representations are all distinct, which is a feature easily missed in the case of $\mathrm{SU}(2)$ owing to the fact that the anti-fundamental representation of $\mathrm{SU}(2)$ is unitarily equivalent to the fundamental. This is, however, not true for $N \ge 3$.}

Other representations can be generated by repeated application of the above. A generic represetation obtained from the fusion of only fundamental anyons consists of multi-index tensors of the following type
\begin{equation}
    T^{(i_1\ldots i_p), (j_1\ldots j_q), \ldots, (k_1\ldots k_r)}, \qquad r \le \ldots \le q \le p
\end{equation}
with the following symmetry properties: 1) all indices of a given group denoted by a Latin letter, e.g.\ $(j_1,\ldots,j_q)$, are symmetric under exchange, and 2) indices with the same numerical subscript are antisymmetric under exchange, e.g.\ $(i_2,j_2,\ldots,k_2)$. These representation are conveniently described by Young tableaux with a box representing an index, rows representing symmetric index groups, and columns representing antisymmetric indices. The number of rows must be less than or equal to $N-1$, since the maximally anti-symmetric combination is trivial. Of these, we will use special labels for a distinct class of representations: $S_n$ for $n$-index symmetric representations and $A_n$ for $n$-index antisymmetric ones. We also note that the anti-fundamental $\overline{F}$ is the same as the anti-symmetric representation $A_{N-1}$.

Not all irreducible representations correspond to anyons. The level $k$ of the Chern-Simons theory puts an upper bound on the number of symmetric indices in the first (longest) row of the Young tableau. In particular, we require
\begin{equation}
    p \le k
\end{equation}
making the number of anyons indeed finite. This means that $S_k$ is the maximally symmetric representation.

The spin-enriched $\mathrm{SU}(N)_k$ theory has an additional ``fermionic'' sector obtained from fusing each allowed anyon with the local fermion $c$\footnote{The spin-enrichment is not needed for odd $k$ since that already requires a spin structure to be defined, but the anyon content is still the same.}. For instance this generates the fermionic cousin $F_c$ of the fundamental representation. Fusion rules then get modified to be consistent with fermion parity. The $F_c$ anyon is what is represented by a quantum of $\psi$ in the fermionic Chern-Simons matter theory, making the maximally antisymmetric representation $A_N$ a local fermion for odd $N$ and a local boson for even $N$. A $\phi$ quantum in the bosonic Chern-Simons matter theory similarly creates an $F$ anyon.


\subsection{'t Hooft large $N$ limit}

The actions in Eq.\ \eqref{eq:CSM_actions_app} are generally strongly coupled and require approximations to simplify. One such approximation is provided by a double-scaling large $N$ limit:
\begin{equation}
    N,k \rightarrow \infty, \qquad \lambda \equiv \frac{N}{\kappa} = \mathrm{const.}, \qquad \kappa = k + \sgn(k) N
\end{equation}
where $\lambda \in (0,1)$ is called the 't Hooft coupling. This limit makes the theory solvable through Schwinger-Dyson equations. The anyon content of the Chern-Simons matter theory limits to that of representations of $\mathrm{SU}(\infty)$ in this limit, owing to the limit $k$ on the width of the Young tableaux limiting to infinity.

That the large $N$ limit of the CSF theory has Fermi-liquid-like behaviour at finite chemical potential was first demonstrated in Ref.~\cite{Geracie2015-ya}. This result was extended to the CSB theory in Refs.~\cite{Minwalla:2020ysu, Minwalla2022-pg}, providing further evidence for the duality between the CSF and CSB theories in the 't Hooft limit under and exchange of $N$ and $k$\footnote{For local observables $\mathrm{SU}(N)_k$ and $\mathrm{U}(N)_k$ are identical in the large $N$ limit.}. They also conjectured a finite $N$ and $k$ expression for the partition function that is a special case of Eq.~\eqref{eq:kin_part_fn_defn_sec3}, defined through what they referred to the \textit{quantum singlet condition}, which is special case of the fusion singlet condition in this paper.


\subsection{Occupation sequences and partition functions}

As mentioned in Section~\ref{sec:Examples_sec5}, the occupation sequences that reproduce the CSF and CSB partition functions of Ref.~\cite{Minwalla2022-pg} are the following
\begin{equation}
    \begin{split}
        \ell_\mathrm{CSF} &= (1, F_c, A_2, \ldots, A_N)\\
        \ell_\mathrm{CSB} &= (1, F, S_2, \ldots, S_k)
    \end{split}
\end{equation}
The quantum dimensions of the symmetric and anti-symmetric representations are given by \emph{q-deformed binomial coefficients}
\begin{equation}
    d_{S_n} = \binom{N+n-1}{n}_q, \qquad d_{A_n} = \binom{N}{n-1}_q, \qquad q = e^{2\pi i/ \kappa}
\end{equation}
which are defined as follows
\begin{equation}
    [n]_q \equiv \frac{q^{n/2}-q^{-n/2}}{q^{1/2}-q^{-1/2}}, \qquad \binom{n}{r}_q \equiv \frac{\Gamma([n]_q + 1)}{ \Gamma([n-r]_q + 1) \Gamma([r]_q + 1) }
\end{equation}
The single-energy-level partition functions then take the following form
\begin{equation}
    \begin{split}
        z_\mathrm{CSF}(\epsilon) &= \sum_{n=0}^N \binom{N}{n}_q y^n = \prod_{j=-(N-1)/2}^{(N-1)/2} \left( 1 + q^j y \right)\\
        z_\mathrm{CSB}(\epsilon) &= \sum_{n=0}^k \binom{N+n-1}{n}_q y^n = \prod_{j=-(N-1)/2}^{(N-1)/2} \frac{1}{\left( 1 - q^j y \right)}\Bigg|_{\text{truncated at }k}
    \end{split}
\end{equation}
where $y=e^{-\beta(\epsilon-\mu)}$ and by ``truncated at $k$'' we mean the geomtric series expansion of $(1-x)^{-1}$ being truncated at $x^k$. The resulting distribution functions take the following form
\begin{equation}
    \begin{split}
        n_\mathrm{CSF}(\epsilon) &= \sum_{j=-(N-1)/2}^{(N-1)/2} \frac{1}{q^{-j} y^{-1} + 1}\\
        n_\mathrm{CSB}(\epsilon) &= \sum_{j=-(N-1)/2}^{(N-1)/2} \frac{1}{q^{-j} y^{-1} - 1} - \frac{\kappa}{y^\kappa - 1}
    \end{split}
\end{equation}
These are identical to Eq. (4.27) in Ref.~\cite{Minwalla2022-pg}.


\subsection{Odds and ends}

Let us briefly comment on some of the issues with this particular sequence construction for Chern-Simons matter theories. Starting with the theory with fermionic matter first, the sequence $\ell_\mathrm{CSF}$ ends at the maximal antisymmetric representation $A_N$ which is indeed a local particle. Depending on whether $N$ is even or odd, this entry might be the local boson $1$ or the local fermion $c$ repsectively. If $N$ is odd, we genuinely have a Fermi-liquid-like state where the electron delocalizes into anyons. However if $N$ is even, the gas is more akin to a correlated insulator where ``Pauli exclusion'' for the local boson $A_N$ is implemented by energetic constraints such as a ``hard-core'' repulsion, which seem to be absent in the microscopic action. The occupation sequence formulation cannot determine whether this is the correct phase at finite $N$ and $k$, but the lack of the required energetic constraints in the microscopic theory makes it more likely that the repeated sequence
\begin{equation}
    \ell'_\mathrm{CSF} = (1, F, A_2, \ldots, A_N = 1, F, A_2, \ldots, A_N = 1, F, \ldots)
\end{equation}
describes the correct finite temperature physics -- that of a superfluid. We thus expect that the large $N$ expansion is unstable to $1/N$ corrections for even $N$. The possibility of a superfluid instability still exists for odd $N$, and might be energetically favoured.

Moving on to the bosonic Chern-Simons matter theory, we have truncated the occupation sequence illegally at $S_k$, which is not a local particle. It is in fact a generator of a $\mathbb{Z}_N$ one-form symmetry with spin $(N-1)/2N$. Even in the large $N$ limit, this remains a non-trivial abelian anyon, but the fact that $k=\infty$ in the strict limit makes this sequence legal. We suspect that subleading corrections in the $1/N$ expansion are important to resolve this, but our sequence formalism provides a possible resolution by simply completing the sequence until $\ell_\mathrm{CSB}$ ends in a local particle.


\section{Relation to Haldane's fractional exclusion principle}\label{app:Haldane_stats_appF}

\subsection{Review of fractional exclusion statistics}

The fractional exclusion statistics of~\cite{Ramanathan1992,Wu1994-dg,Nayak1994,Murthy:1994yew,Ha1994,Isakov:1994zz,deVeigy1994,Polychronakos1995-kd,Isakov:1996aj,Schoutens:1997xh,Murthy1999-rf,Ardonne2001-wx,Ouvry2009-ii,Xiong:2022mll} rely on a simple starting point: that the number of states available for a given anyon to occupy decreases uniformly with the number of anyons already occupying the rest of the Hilbert space. In equations,
\begin{equation}
    \Delta \mathfrak{D} (n) = - g \Delta n
\end{equation}
where $\mathfrak{D}(n)$ is the dimension of the available Hilbert space for the addition of one anyon to a state consisting of $n$ anyons, and $g$, the \textit{statistical factor}, is a number ranging from 0 to 1 and independent of $n$. The extremal values of $g=0$ and $g=1$ correspond to bosons and fermions respectively. A fractional value of $g$ corresponds to anyons. One can generalize this to multi-anyon mixtures by adding anyon indices to everything
\begin{equation}
    \Delta \mathfrak{D}_a = - \sum_b g_{ab} \Delta n_b
\end{equation}
making $g_{ab}$ a \textit{statistical matrix} instead. Below we will demonstrate how to reproduce this from the occupation sequence formalism with one simple generalization: the statistical factor can now depend on $n$:
\begin{equation}
    \Delta \mathfrak{D} = - g(n) \Delta n
\end{equation}
This $n$-dependent statistical factor still admits an equivalent description with a constant statistical matrix $g_{ij}$ whose indices then count the occupation number and whose components are 0 unless $j=i$ or $j=i-1$. One can use these two descriptions interchangeably by promoting the function to a matrix.

Suppose we define $\mathcal{H}_n$ as the available Hilbert space for the $n$th particle in the background on $n-1$ particles, upon which we get
\begin{equation}
    \mathfrak{D}(n) \equiv \dim \mathcal{H}_n
\end{equation}
Whenever we dope one more particle to the system, the dimension is reduced as 
\begin{equation}
    \begin{split}
        \mathfrak{D}(n) &= \mathfrak{D}(n-1) - g(n-1)\\
        &= \mathfrak{D}(1) - g(1) - g(2) - \ldots g(n-1)
    \end{split}
\end{equation}
This process ends at the value $n_\mathrm{max} = N$ whenever we get $\mathfrak{D}(N) \leq g(N)$ and thus no more doping is allowed. Therefore, the $\mathfrak{D}(n)$'s should always be positive as long as $1\leq n \leq N$. For example, the fermion has $N=\mathfrak{D}(1)$ while the boson has $N=\infty$. Since fractional exclusion statistics have so far required that $g(n)=g$ be fractional, strictly positive, and independent of $n$, the sequence of $\mathfrak{D}(n)$'s always truncates, precluding the possibility of Bose condensation by hand.

We can then derive the degeneracy, i.e., the number of allowed $n$-particle states, as
\begin{equation}\label{eq:stat_weight}
    W_n [g]= \begin{pmatrix}
        \mathfrak{D}(n) + n -1  \\ n
    \end{pmatrix}=
    \begin{pmatrix}
        \mathfrak{D}(1) + n -1 - \sum_{k=1}^{n-1}g(k) \\ n
    \end{pmatrix}
\end{equation}
We assume $W_0 [g]=1$ below. The partition function per energy level of the gas can then be written down directly:
\begin{equation}\label{eq:Z_haldane_stat}
    z(\epsilon) =\sum_{n=0}^N W_n[g]\  e^{-n \beta  (\epsilon-\mu)}.
\end{equation}
This has the same form as the partition function derived in Section~\ref{sec:distr_fn_occupation_sequence_sec4}, specifically Eq.~\eqref{eq:single_level_part_fn}, up to the following replacement:
\begin{equation}
    \begin{split}
        |\ell| &\longleftrightarrow N,\\
        d_{a_n}  &\longleftrightarrow W_n[g],
    \end{split}
\end{equation}
This dictionary makes it immediately clear why partition functions derived from fractional exclusion statistics with constant $g(n)$ can never give rise to superfluids, since we always have a finite $N=|\ell|$. This translation makes it clear why we require the statistical factor $g$ to depend on the occupation of the state. For a constant factor $g$, given a sequence of quantum dimensions $d_{a_n}$, we would need to solve the following set of $N$ equations
\begin{equation}
    d_{a_n} = \binom{\mathfrak{D}(1)+(1-g)(n-1)}{n}, \qquad n = 1, \ldots, N
\end{equation}
which in general has no solution except for specific occupation sequences. On the other hand, if we have $N$ independent variables $g(n)$, the set of equations
\begin{equation}
    d_{a_n} = \binom{\mathfrak{D}(1)+n-1-\sum_{k=1}^{n-1}g(k)}{n}, \qquad n = 1,\ldots, N
\end{equation}
always has a solution. On the other hand, not all arbitrary values of the statistical factor $g(n)$ admit a description in terms of an occupation sequence, because the occupation sequence is derived from a modular tensor category which imposes many additional constraints on the quantum dimensions coming from fusion and braiding.

In the following we will revisit the examples from Section~\ref{sec:Examples_sec5} and describe the corresponding statistical factor.


\subsection{Statistical factor from occupation sequences}


\subsubsection{Fermionic Laughlin states}

Recall again that the occupation sequence for fermionic Laughlin states at filling $1/k$ for odd $k$ was given by
\begin{equation}
    \ell_k = (1, a_e, a^2, \ldots, e)
\end{equation}
where $e$ is the electron and $a_e$ is the quasiparticle of charge $e/k$. Since this is an abelian theory, all quantum dimensions are $1$, and we need to solve the following equations
\begin{equation}
    1 = \begin{pmatrix} \mathfrak{D}(n)+n-1 \\ n \end{pmatrix}, \qquad n = 1, \ldots k,
\end{equation}
the solution to which is presented in Table~\ref{tab:laughlin_Haldane_appE}.

\begin{table}[h]
    \centering
    \begin{tabular}{c|ccccc} 
        $n$ &  $1$ & $2$ & $\ldots$ & $k-1$ & $k$   \\ \hline
        $\mathfrak{D}(n)$ & $1$ & $1$ & $\ldots$ & $1$ & $1$ \\ \hline
        $g(n)$ & $0$ & $0$ & $\ldots$ & $0$ & $1$  \\ \hline
        $W_n$ & $1$ & $1$ & $\ldots$ & $1$ & $1$ 
    \end{tabular}
    \caption{Fractional exclusion statistics for fermionic Laughlin states.}
    \label{tab:laughlin_Haldane_appE}
\end{table}

Note that the fact that $g(n) = 0$ until maximum occupancy is achieved is in stark constrast with the more historic approach of describing fermion Laughlin states with the statistical factor $g=1/k$ which slowly reduces the available Hilbert space until it is all gone. Furthermore, the values of the statistical factor that we obtain are also integers, making it seem more reasonable to refer to $\mathfrak{D}(n)$ as dimensions of Hilbert spaces. This, however, is an artifact of abelian theories where the quantum dimensions are always integers (specifically, 1), allowing for integer solutions to the statistical factor.


\subsubsection{Fermionic Fibonacci anyons}

We next consider a non-abelian example: the fermionic Fibonacci gas, whose occupation sequence is given by
\begin{equation}
    \ell_{\mathrm{Fib}_\psi} = (1, \psi_\psi, \tau, \psi)
\end{equation}
The system of equations to solve becomes the following:
\begin{equation}
    d_\tau = \binom{\mathfrak{D}(1)}{1} = \binom{\mathfrak{D}(2) + 1}{2},  \qquad
    1 = \binom{\mathfrak{D}(3) + 2}{3}.
\end{equation}
The numerical values of these are presented in Table~\ref{tab:fib_fermi_Haldane_appE}.
\begin{table}[h]
    \centering
    \begin{tabular}{c|ccc}
        $n$ &  $1$ & $2$ & $3$\\ \hline
        $\mathfrak{D}(n)$ & $d_\tau=1.618$ & $1.367$ & $1$ \\ \hline
        $g(n)$ & $0.251$ & $0.367$ & $d_\tau^{-1} = 0.618$  \\ \hline
        $W_n$ & $d_\tau$ & $d_\tau$ & $1$ 
    \end{tabular}
    \caption{Fractional exclusion statistics for the fermionic Fibonacci gas.}
    \label{tab:fib_fermi_Haldane_appE}
\end{table}


\subsubsection{Bosonic Fibonacci anyons}

Let us now demonstrate how making the statistical factor dependent upon the occupation number allows superfluid partition functions using the bosonic Fibonacci gas as an example. The relevant sequence is
\begin{equation}
    \ell_\mathrm{Fib} = (1,\tau,1,\tau,\ldots)
\end{equation}
which means we must solve the following equations
\begin{equation}
    \binom{\mathfrak{D}(n) + n - 1}{n} = \begin{cases}
        1, \qquad &n \text{ even}\\
        d_\tau, \qquad &n \text{ odd}
    \end{cases}
\end{equation}
The solution for even $n$ is straightforward: $\mathfrak{D}(n) = 1$. The odd $n$ equations are solve numerically and the results are presented in Table \ref{tab:fib_bose_Haldane_appE}
\begin{table}[h]
    \centering
\begin{tabular}{c|cccccccc} 
$n$ &  $1$ & $2$ & $3$ & $4$ & $5$ & $\ldots$ \\ \hline
 $\mathfrak{D}(n)$ & $d_\tau = 1.618$ & $1$ & $1.289$ & $1$ & $1.225$ & $\ldots$\\ \hline
 $g(n)$ & $d_\tau^{-1}=0.618$ & $-0.289$ & $-0.289$ & $-0.225$ & $0.225$ & $\ldots$ \\ \hline
 $W_n $ & $d_\tau$ & $1$ & $d_\tau$ & $1$ & $ d_\tau$ & $\ldots$
\end{tabular}
    \caption{Fractional exclusion statistics for the bosonic Fibonacci gas.}
    \label{tab:fib_bose_Haldane_appE}
\end{table}
Unlike a regular Bose gas for which $g(n) = 0$ for all values of $n$ we find that for the non-abelian superfluid we find that $g(n)$ alternates in sign with the magnitude decreasing with the period of the sequence, i.e., every two entries, eventually asymptoting to the Bose gas value $0$.

\end{appendix}

\nocite{apsrev41Control}
\bibliographystyle{apsrev4-1}
\bibliography{anyon_thermo}

\end{document}